\documentclass[%
  reprint,
  twocolumn,
 showpacs,
 showkeys,
 preprintnumbers,
 nofootinbib,
 amsmath,amssymb,
 aps,
 pra,
  longbibliography,
 ]{revtex4-1}

\usepackage[dvipsnames]{xcolor}

\usepackage{mathptmx}

\usepackage{amssymb,amsthm,amsmath}

\usepackage{tikz}
\usetikzlibrary{calc,decorations.pathreplacing,decorations.markings,positioning,shapes}

\usepackage[breaklinks=true,colorlinks=true,anchorcolor=blue,citecolor=blue,filecolor=blue,menucolor=blue,pagecolor=blue,urlcolor=blue,linkcolor=blue]{hyperref}
\usepackage{graphicx}
\usepackage{url}

\begin{document}

\title{Extensions of Hardy-type true-implies-false gadgets to classically obtain indistinguishability}

\author{Karl Svozil}
\email{svozil@tuwien.ac.at}
\homepage{http://tph.tuwien.ac.at/~svozil}

\affiliation{Institute for Theoretical Physics,
TU Wien,
\\
Wiedner Hauptstrasse 8-10/136,
1040 Vienna,  Austria}

\date{\today}

\begin{abstract}
In quantum logical terms, Hardy-type arguments can be uniformly presented and extended as collections of intertwined contexts and their observables. If interpreted classically those structures serve as graph-theoretic ``gadgets'' that enforce correlations on the respective preselected and postselected observable terminal points. The method allows the generalization and extension to other types of relational properties, in particular, to novel joint properties predicting classical equality of quantum mechanically distinct observables. It also facilitates finding faithful orthogonal representations of quantum observables.
\end{abstract}


\maketitle

\setcounter{MaxMatrixCols}{20}

\section{Certification of nonclassicality}

When it comes to certifying nonclassical observance of quantized systems at least three types of approaches have been suggested:
(i) Bell-type theorems, related to Boole's ``conditions of possible experience'' for observables in disjoint~\cite{froissart-81,cirelson,Pit-94}
or intertwined~\cite{svozil-2001-cesena,Klyachko-2008,cabello:210401,svozil-2017-b} contexts
(also known as maximal collection of compatible observables organized in a Boolean subalgebra),
present empirical evidence involving statistical terms which (due to complementarity) cannot be obtained simultaneously, but
are obtained from sequential ``one term at a time'' measurements: whatever the sampling,
the events contributing to each term need to be temporally (mutually) apart from the other terms.
(ii) Kochen-Specker type theorems~\cite{kochen1,pitowsky:218,cabello-96,PhysRevA.89.032109,2015-AnalyticKS}
are theoretical proofs by contradiction employing finite sets
of intertwined quantum observables~\cite{gleason}
which have no classical interpretation in terms of two-valued (truth) assignments.
Indirect empirical corroborations of Kochen-Specker type theorems (and, thereby, quantum contextuality)
amount to violations of local Bell-type classical predictions~\cite{cabello:210401}.
(iii)~A third, statistical, method~\cite{2018-minimalYIYS,svozil-2017-b}
is based on preselected (prepared) quantum states
which are sequentially measured or postselected in terms of
suitably chosen quantum observables: prepare one state, measure another.
Thereby, pre- and postselection (and their respective observables) are
imagined to be logically connected by suitable finite collections of
hypothetical counterfactual~\cite{specker-60} intertwined contexts, with their choice being motivated by their predictive capacities
and yet remaining arbitrary~\cite{svozil-2020-c}.
In particular, particles are prepared in such states and observable properties,
and are logically connected to other observables in certain ways,
such that their (non)occurrence is classically mandatory, but quantum mechanically unrestricted,
and occasionally violates the respective classical predictions.
By choosing different patterns of connection, this method could be
strengthened to the point that any singular outcome contradicts the respective classical predictions.
With this (counterfactual) adaptive modification, any such stochastic argument turns definite~\cite{svozil-2020-c}.

Common to all these cases is their reliance on complementary counterfactuals~\cite{specker-60}
because they suppose the simultaneous existence of more than one context:
Except for common observables of intertwining~\cite{gleason} contexts,
all observables in any such context are mutually complementary to all observables in a different context,
and there is no physically feasible way of simultaneously measuring them.

For the sake of obtaining discrepancies between classical and quantum predictions
these conglomerates of contexts, and thereby the quantum observables they consist of,
are interpreted ``as if'' they could have a classical interpretation.
That is, a classical interpretation is forced upon such collections of (intertwining) observables.
In quantum logic, a classical interpretation amounts to a two-valued (also known as $0-1$, false-true) state
(frame function~\cite{gleason}),
which is context-independent; that is, its value on observables does not depend on the particular context
(also known as maximal observable~\cite[\S~84, p.~172,173]{halmos-vs}, orthonormal basis) in which it occurs.

Kochen-Specker type theorems employ configurations of intertwining contexts
for which these classical interpretations (in terms of two-valued measures interpreted as dichotomic observables) fail:
There does not exist any consistent classical interpretation for these conglomerates of contexts and the observables they hold.

Type-(iii) configurations exhibiting finite collections of observables (in intertwining contexts)
may still allow classical interpretations,
but the predictions based upon them statistically directly contradict the quantum predictions
and without the need of type-(i) inequalities~\cite{Hardy-92}.
Indeed, Kochen and Specker used the latter~\cite{kochen2} and constructed the former, stronger result~\cite{kochen1}
without explicitly mentioning the empirical opportunities of such configurations.
Appendix~B on pages~64-65 in \cite{Belinfante-73},
pp.~588--589 in \cite{stairs83}
provide early discussions~\cite[Section~5.5.3]{svozil-2017-b} of type-(iii) nonclassicality.

In what follows we
shall extend a particular type~(iii) instance involving two two-state particles
proposed by Hardy~\cite{Hardy-92,Hardy-93} to configurations
that contain distinct quantum observables that cannot be ``resolved'' or distinguished classically.
En route we shall study configurations~\cite{2018-minimalYIYS} of contexts enforcing classical predictions which are the ``inverse''
of the original relational properties resulting from Hardy-type arguments.
The latter are often synonymously referred to as
``Hardy's theorem''~\cite{Goldstein-1994,cabello-96},
``Hardy's proof''~\cite{cabello-97-nhvp,Cabello-2013-Hardylike,2018-minimalYIYS,Xu-Chen-Guehnw-2020}, or
``Hardy's paradox''~\cite{Badziag-2011,Cabello-2013-HP}.
``Hardy's wonderful trick''~\cite{mermin-1995}, also called
``Hardy's beautiful example''~\cite[Section~23.5, p.~589ff]{penrose:2004}, and
attempts to make it accessible to a wider audience abound~\cite{Mermin1994-Hardy,Kwiat-Hardy-2000}.
Nevertheless, a detailed account in terms of the quantum logical structure of the counterfactual argument
results in a better comprehension of the resources and assumptions involved;
in particular, when it comes to related proposals.
Such an account suggests extensions to other type-(iii) configurations
with different relational properties, and
also allows empirical predictions
by yielding a systematic way of finding quantum realizations,
in particular, in regard to desiderata such as (in)distinguishability
of the associated quantized entities.

\section{Hypergraph nomenclature}

The counterfactual arguments (ii) and (iii) mentioned earlier
can be depicted structurally  transparent by the use of
hypergraphs~\cite{kalmbach-83,svozil-tkadlec,2018-minimalYIYS,svozil-2017-b}
introduced by Greechie, drawing contexts as smooth lines.
[In what follows we shall use the following terms synonymously, thereby having in mind the different areas in which they occur:
context, maximal observable, orthonormal basis, block, (Boolean) subalgebra, (maximal) clique, and complete graph.]
In particular,
Greechie has suggested to (amendments are indicated by square brackets)~\cite[p.~120]{greechie:71}
{\em ``[$\ldots$]
present [$\ldots$] lattices as unions of [contexts]
intertwined or pasted together in some fashion
[$\ldots$]
by replacing, for example, the $2^n$ elements in the Hasse diagram of the power set
of an $n$-element set with the [context] complete graph [$K_n$] on $n$ elements.
The reduction in numbers of elements is considerable but the number of remaining `links'
or `lines' is still too cumbersome for our purposes.
We replace the [context] complete graph on $n$ elements by a single smooth curve (usually a straight line)
containing $n$ distinguished points. Thus we replace $n(n + 1)/2$ ``links'' with a single smooth curve.
This representation is propitious and uncomplicated provided that
the intersection of any pair of blocks contains at most one atom.''}

In what follows we shall refer to such a general representation of observables as the {(orthogonality) hypergraph}~\cite{kalmbach-83}.
The term should be understood in the broadest possible consistent sense.
Most of our arguments will be in four-dimensional state space.
An exception will be our mentioning the
gadget\footnote{A clarification with regards
to the use of the technical term ``gadget'' seems in order: this denomination is frequently used in graph theory to indicate
``useful subgraphs''. It is not meant to be polemic.}~\cite{tutte_1954,SZABO2009436,Ramanathan-18},
called  ``K\"afer'' (German for ``bug'')  by  Specker,
which was introduced in 1965~\cite{kochen2}
and used in the Kochen-Specker proof~\cite{kochen1},
serving as a true-implies-false configuration.
The Specker bug is the three-dimensional analog of the four-dimensional Hardy configuration~\cite{2018-minimalYIYS}.
Its hypergraph is depicted in Fig.~\ref{2020-hardy-fig1}(b).

We shall concentrate on orthogonality hypergraphs which are pasting~\cite{Greechie1968} constructions~\cite[Chapter~2]{greechie-66-PhD}
of a homogeneous  single type
of contexts $K_n$,
where the (maximal) clique number $n$ is fixed.
(Note that other authors use related definitions for Greechie diagrams~\cite{Mckay2000} and McKay-Megill-Pavicic (MMP) diagrams~\cite{Pavicic-2005}.)
If interpreted as representing some configuration of mutually orthogonal vectors,
the (maximal) clique number $n$ equals the dimension of the Hilbert space.

\section{Faithful orthogonal representations of (hyper)graphs}

In what follows, ket vectors, which are usually represented by column vectors, will be represented by the respective transposed row vectors.
In all our examples, hypergraphs have a
faithful orthogonal representation~\cite{lovasz-79,lovasz-89,Grtschel1993,Portillo-2015}
in terms of vectors which are mutually orthogonal within (maximal) cliques or contexts.
The phrases faithful orthogonal representation, coordinatization~\cite{Pavii2018} or vector encoding will be used synonymously.

When drawing hypergraphs,
some of the atomic propositions will be omitted (or only drawn lightly) if they are not essential to the argument.
In particular, in three and four dimensions, given two orthogonal (in general noncollinear) vectors it is always possible to ``complete'' this
partially defined context by a Gram-Schmidt  process~\cite{svozil-tkadlec,Pavi_i__2019}.
Indeed, given two (orthogonal) noncollinear vectors, in three dimensions the span of the ``missing'' vector
is uniquely determined by the span of the cross product of those two vectors.
(A generalized cross product of $n-1$  vectors
in $n$-dimensional space can be written as a determinant; that is, in the form of a Levi-Civita symbol.)
This ``lack of freedom'' is in one dimension;
in particular, whenever the missing vector is collinear to some vector occurring
in the faithful orthogonal representation of the incomplete hypergraph that one is attempting to complete.
The most elementary such counterexample is a triangular hypergraph with three cyclically connected contexts $\{\{1,2,3\},\{3,4,5\},\{5,6,1\}\}$.
Consider any incomplete faithful orthogonal of its intertwining atoms such as
$1=\begin{pmatrix}0,0,1\end{pmatrix}$,
$3=\begin{pmatrix}0,1,0\end{pmatrix}$,
$5=\begin{pmatrix}1,0,0\end{pmatrix}$: any conceivable completion fails because the missing vectors would result
in duplicities in the faithful orthogonal representation, that is, in $2=5$,  $4=1$, and $6=3$.

Nevertheless, in four dimensions, given at least two (orthogonal) noncollinear vectors,
the two-dimensional orthogonal subspace is spanned by a continuity of (e.g. rotated) bases.
Therefore, in such a case there is always ``enough room for breathing''; that is,
for accommodating the basis vectors and thereby transforming them if necessary
such that any hypergraph can be properly completed without duplicities.
I encourage the reader to try to find a faithful orthogonal representation of the cyclic triangular shaped hypergraph
$\{\{1, .  .  ,4\}$,
$\{4, .  . ,7\}$,
$\{7, .  . ,1\}\}$ in four dimensions.

Whether or not such faithful orthogonal representations can be given in terms of
decomposable or indecomposable vectors associated with factorizable or entangled states is an entirely different issue.
In four dimensions a careful analysis~\cite{havlicek-svozil-2020-dec}
yields a no-go theorem for four-dimensional coordinatizations of the triangle hypergraph
by allowing a maximal number 7 of 9 decomposable vectors.

In general and for arbitrary dimensions,
as long as there are two or more ``free'' (without any strings and intertwining contexts attached)
vectors per context missing from a faithful orthogonal representation of a hypergraph, its completion is always possible.
Stated differently, any faithful orthogonal representation of an incomplete hypergraph can be straightforwardly extended
(without reshuffling of vector components) to a faithful orthogonal representation
in a completed hypergraph (eg, by a Gram-Schmidt process)
if at least two or more nonintertwining vectors per context in that hypergraph are missing.
Indeed, one may even drop an already existing vector ``blocking'' a faithful orthogonal representation of an entire (hyper)graph
if the associated atom is not intertwining in two or more contexts,
and if the new freedom facilitates continuous bases
instead of a single vector whose addition may result in duplicities through collinear vectors (we shall mention such an instance later).

In the case of two or more ``missing'' vectors, any completion involves a two- or higher-dimensional subspace.
Any such subspace $\mathbb{R}^{k\ge n-2}$ or $\mathbb{C}^{k\ge n-2}$  of the $n$-dimensional continua $\mathbb{R}^{n}$ or $\mathbb{C}^{n}$
is spanned by a continuity of (orthogonal) bases.
A typical example is an incomplete faithful orthogonal representation of a basis of $\mathbb{R}^4$ rotated into a form
$\left\{\begin{pmatrix}1,0,0,0\end{pmatrix},\begin{pmatrix}0,1,0,0\end{pmatrix}\right\}$.
Its completion is then given by the continuity of bases
$\left\{
\begin{pmatrix}0,0,\cos \theta ,\sin \theta \end{pmatrix},
\begin{pmatrix}0,0,-\sin \theta , \cos \theta \end{pmatrix}
\right\}$,
with $0\le \theta < \pi$.

A completion should even be possible if one merely allows sets of bases which are denumerable---or even finitely
but ``sufficiently'' many bases with respect to the hypergraph encoded.
From this viewpoint four dimensions offer a much wider variety of completions as compared to the three-dimensional case---indeed, the difference results from the abundance offered by a continuum of bases versus a single vector spanning the respective subspaces,
a fact which is very convenient for all kinds of constructions.
However, as has been mentioned earlier, the completion of coordinatizations of hypergraphs by
(in)decomposable vectors---in particular, if one desires to maintain (non)decomposability---is an altogether different issue~\cite{havlicek-svozil-2020-dec}.

This possibility to complete incomplete contexts is also the reason why practically all papers introducing and reviewing Hardy's configuration operate
not with the complete eight contexts including 21 atomic vertices, but merely with the nine vectors or vertices in which those eight contexts intertwine.
Nevertheless, for tasks such as determining whether a particular configuration of observables supports or does not allow a classical two-valued
state, as well as for determining the set of two-valued states and their properties (e.g., separable, unital),
the nonintertwining atomic propositions matter.

\section{Quantum logical formulation of Hardy's argument}

For the sake of being able to delineate Hardy's rather involved original derivation~\cite{Hardy-93},
let us stick to his nomenclature as much as possible.
We shall, however, drop the particle index as it is redundant.
So, for instance, Hardy's
$
\vert + \rangle_1
\vert + \rangle_2
$
will be written as
$
\vert + \rangle
\vert + \rangle
=
\vert ++ \rangle
$.
Later, we shall be very explicit and identify the respective entities in terms of Hardy's Ansatz,
but let us study Hardy's schematics in some generality first:

\begin{enumerate}
\item[(i)]
It begins with a specific entangled state of two two-state particles $\vert \Psi \rangle$.

\item[(ii)]
Then the argument suggests measuring two dichotomic (i.e. two-valued) observables
$\hat{U}$ (exclusive) or $\hat{D}$ on each one of the two particles.
This results in four measurement configurations
$\hat{U} \otimes \hat{U}$,
$\hat{U} \otimes \hat{D}$
$\hat{D} \otimes \hat{U}$
$\hat{D} \otimes \hat{D}$---that is, effectively, the two-particle observable
$\hat{U} \otimes \hat{D}$
is measured ``in Einstein-Podolsky-Rosen (EPR) terms of''
$\hat{U} \otimes \mathbb{I}_2$
and
$\mathbb{I}_2 \otimes \hat{D}$.

\item[(iii)]
As both of these dichotomic observables $\hat{U}$ and $\hat{D}$
have two possible outcomes called
$u$ and $v$ for $\hat{U}$
and
$c$ and $d$ for $\hat{D}$, respectively,
there are $2^2 \times 2^2 = 2^4 = 16$ different outcomes that are denoted by the ordered pairs
$uu$,
$uv$,
$uc$,
$ud$,
$vu$,
$vv$,
$vc$,
$vd$,
$cu$,
$cv$,
$cc$,
$cd$,
$du$,
$dv$,
$dc$, and
$dd$.

\item[(iv)]
From these 16 outcomes 5 groups of
(incomplete if not all atoms or vertices are specified---yet, as discussed earlier, a completion is straightforward if desired)
contexts which consist of simultaneously measurable and mutually exclusive observables can be formed, namely
$\{dd, .  . ,  cv  \}$,
$\{dd, .  . ,  vc  \}$,
$\{cv,vu,uu,dv\}$,
$\{vc,uv,uu,vd\}$, and
$\{vu, .  . ,uv\}$.

\item[(v)]
Finally,  this collection of five contexts are ``bundled  with'' or ``tied to'' the (projection)
observable corresponding to the original
entangled state $\vert \Psi \rangle$ introduced in (i) by the three (incomplete) contexts
$\{vd, .  . ,\Psi \}$,
$\{uu, .  . ,\Psi \}$, and
$\{dv, .  . ,\Psi \}$.
\end{enumerate}

\begin{figure}
\begin{center}
\begin{tabular}{ c c c }
\begin{tikzpicture}  [scale=0.5]

\tikzstyle{every path}=[line width=1pt]

\newdimen\ms
\ms=0.1cm
\tikzstyle{s1}=[color=red,rectangle,inner sep=3.5]
\tikzstyle{c3}=[circle,inner sep={\ms/8},minimum size=4*\ms]
\tikzstyle{c2}=[circle,inner sep={\ms/8},minimum size=3*\ms]
\tikzstyle{c1}=[circle,inner sep={\ms/8},minimum size=2*\ms]
\tikzstyle{cs1}=[circle,inner sep={\ms/8},minimum size=1*\ms]


\coordinate (psi) at (2,0);
\coordinate (vd) at (0,2);
\coordinate (dv) at (4,2);
\coordinate (uu) at (2,4);
\coordinate (vu) at (1,5);
\coordinate (uv) at (3,5);
\coordinate (cv) at (0,6);
\coordinate (vc) at (4,6);
\coordinate (dd) at (2,8);


\draw [color=orange] (psi) -- (vd)  coordinate[cs1,fill=white,draw=gray,pos=0.33,label=below left:{\scriptsize \color{gray}2}] (2)  coordinate[cs1,fill=white,draw=gray,pos=0.66,label=below left:{\scriptsize \color{gray}3}] (3);
\draw [color=blue] (psi) -- (uu)  coordinate[cs1,fill=white,draw=gray,pos=0.33,label={[right,xshift=0.2mm]:{\scriptsize \color{gray}20}}] (20)  coordinate[cs1,fill=white,draw=gray,pos=0.66,label={[right,xshift=0.2mm]:{\scriptsize \color{gray}21}}] (21);
\draw [color=red] (psi) -- (dv) coordinate[cs1,fill=white,draw=gray,pos=0.33,label=below right:{\scriptsize \color{gray}17}]  (17) coordinate[cs1,fill=white,draw=gray,pos=0.66,label=below right:{\scriptsize \color{gray}16}] (16);
\draw [color=green] (vd) -- (vc);
\draw [color=gray] (dv) -- (cv);
\draw [color=magenta] (vu) -- (uv) coordinate[cs1,fill=white,draw=gray,pos=0.33,label=above:{\scriptsize \color{gray}18}] (18)  coordinate[cs1,fill=white,draw=gray,pos=0.66,label=above:{\scriptsize \color{gray}19}] (19);
\draw [color=cyan] (cv) -- (dd) coordinate[cs1,fill=white,draw=gray,pos=0.33,label=above left:{\scriptsize \color{gray}8}]  (8) coordinate[cs1,fill=white,draw=gray,pos=0.66,label=above left:{\scriptsize \color{gray}9}] (9);
\draw [color=olive] (vc) -- (dd) coordinate[cs1,fill=white,draw=gray,pos=0.33,label=above right:{\scriptsize \color{gray}12}] (12)  coordinate[cs1,fill=white,draw=gray,pos=0.66,label=above right:{\scriptsize \color{gray}11}] (12);


\draw (psi) coordinate[c3,fill=orange,label=below:$\Psi$];
\draw (psi) coordinate[c2,fill=red];
\draw (psi) coordinate[c1,fill=blue];

\draw (vd) coordinate[c2,fill=orange,label=left:$vd$];
\draw (vd) coordinate[c1,fill=green];

\draw (dv) coordinate[c2,fill=red,label=right:$dv$];
\draw (dv) coordinate[c1,fill=gray];

\draw (uu) coordinate[c3,fill=gray,label=right:$uu$];
\draw (uu) coordinate[c2,fill=green];
\draw (uu) coordinate[c1,fill=blue];

\draw (vu) coordinate[c2,fill=gray,label=left:$vu$];
\draw (vu) coordinate[c1,fill=magenta];

\draw (uv) coordinate[c2,fill=green,label=right:$uv$];
\draw (uv) coordinate[c1,fill=magenta];

\draw (cv) coordinate[c2,fill=gray,label=left:$cv$];
\draw (cv) coordinate[c1,fill=cyan];

\draw (vc) coordinate[c2,fill=green,label=right:$vc$];
\draw (vc) coordinate[c1,fill=olive];

\draw (dd) coordinate[c2,fill=olive,label=above:$dd$];
\draw (dd) coordinate[c1,fill=cyan];

\end{tikzpicture}

&

\begin{tikzpicture}  [scale=1]

\tikzstyle{every path}=[line width=1pt]

\newdimen\ms
\ms=0.1cm
\tikzstyle{s1}=[color=red,rectangle,inner sep=3.5]
\tikzstyle{c3}=[circle,inner sep={\ms/8},minimum size=5*\ms]
\tikzstyle{c2}=[circle,inner sep={\ms/8},minimum size=3*\ms]
\tikzstyle{c1}=[circle,inner sep={\ms/8},minimum size=2*\ms]
\tikzstyle{cs1}=[circle,inner sep={\ms/8},minimum size=1*\ms]


\coordinate (a8) at  (1,2);
\coordinate (a7) at (2,1);
\coordinate (a4) at (0.5,0.5);
\coordinate (a3) at (2,-1);
\coordinate (a1) at (1,-2);
\coordinate (a2) at (0,-1);
\coordinate (a5) at (1.5,0.5);
\coordinate (a6) at (0,1);


\draw [color=orange] (a1) -- (a3)  coordinate[cs1,fill=white,draw=gray,pos=0.5] (2);
\draw [color=blue] (a3) -- (a6);
\draw [color=red] (a6) -- (a8)  coordinate[cs1,fill=white,draw=gray,pos=0.5] (7);
\draw [color=green] (a7) -- (a8)  coordinate[cs1,fill=white,draw=gray,pos=0.5] (3);
\draw [color=gray] (a2) -- (a7);
\draw [color=magenta] (a2) -- (a1)  coordinate[cs1,fill=white,draw=gray,pos=0.5] (10);
\draw [color=cyan] (a4) -- (a5)  coordinate[cs1,fill=white,draw=gray,pos=0.5] (13);


\draw (a1) coordinate[c2,fill=orange,label=below:$a_1$];
\draw (a1) coordinate[c1,fill=magenta];

\draw (a2) coordinate[c2,fill=magenta,label=left:$a_{2}$];
\draw (a2) coordinate[c1,fill=gray];

\draw (a3) coordinate[c2,fill=blue,label=right:$a_3$];
\draw (a3) coordinate[c1,fill=orange];

\draw (a4) coordinate[c2,fill=cyan,label=left:$a_4$];
\draw (a4) coordinate[c1,fill=blue];

\draw (a5) coordinate[c2,fill=gray,label=right:$a_{5}$];
\draw (a5) coordinate[c1,fill=cyan];

\draw (a6) coordinate[c2,fill=red,label=left:$a_6$];
\draw (a6) coordinate[c1,fill=blue];

\draw (a7) coordinate[c2,fill=gray,label=right:$a_7$];
\draw (a7) coordinate[c1,fill=green];

\draw (a8) coordinate[c2,fill=green,label=above:$a_8$];
\draw (a8) coordinate[c1,fill=red];

\end{tikzpicture}
\\
(a) & (b)
\end{tabular}
\end{center}
\caption{\label{2020-hardy-fig1}
Orthogonality hypergraphs of
(a) the Hardy gadget
with 8 contexts and 21 atoms
$\{
 \{dd,8,9,cv\}     $,
$\{dd,11,12,vc\}   $,
$\{cv,vu,uu,dv\}   $,
$\{vc,uv,uu,vd\}   $,
$\{vu,18,19,uv\}   $,
$\{vd,2,3,\Psi \}  $,
$\{uu,20,21,\Psi \}$,
$\{dv,16,17,\Psi \}
\}$;
(b) rendition of the true-implies-false Specker bug gadget
with 7 contexts and 13 atoms
$\{
\{a_8,.,a_6\}  $,
$\{a_8,.,a_7\}  $,
$\{a_6,a_4,a_3\}  $,
$\{a_7,a_5,a_2\}  $,
$\{a_4,.,a_5\}    $,
$\{a_2,.,a_1 \}   $,
$\{a_3,.,a_1 \}
\}$.
Small circles indicate ``auxiliary'' observables which can be choosen freely,
subject to orthogonality constraints: all smooth lines indicate respective contexts representing orthonormal bases.
Larger circles indicate observables common to two or more intertwining contexts.}
\end{figure}
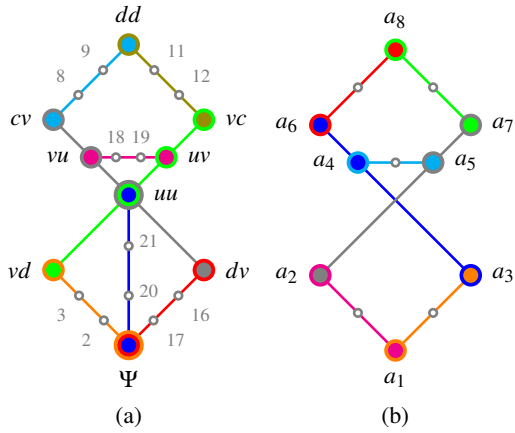

As a result these (incomplete) contexts,
if pasted~\cite{Greechie1968} together at their respective intertwining observables,
result in a collection of eight (incomplete) contexts
\begin{equation}
\begin{split}
\{
\{dd, .  . ,cv\}    = \{dd,8,9,cv\}         ,        \\
\{dd, .  . ,vc\}    = \{dd,11,12,vc\}         ,        \\
\{cv,vu,uu,dv\}     = \{cv,vu,uu,dv\}          ,         \\
\{vc,uv,uu,vd\}     = \{vc,uv,uu,vd\}          ,         \\
\{vu, .  . ,uv\}    = \{vu,18,19,uv\}         ,        \\
\{vd, .  . ,\Psi \} = \{vd,2,3,\Psi \}      ,     \\
\{uu, .  . ,\Psi \} = \{uu,20,21,\Psi \}      ,     \\
\{dv, .  . ,\Psi \} = \{dv,16,17,\Psi \}
\}
\end{split}
\label{2020-hardy-e-hardyset}
\end{equation}
whose orthogonality hypergraph is depicted in Fig.~\ref{2020-hardy-fig1}(a).

\subsection{Classical realization and predictions}

In what follows, we shall prove the following:
\begin{enumerate}

\item[(i)] Hardy's configuration~(\ref{2020-hardy-e-hardyset}) allows a classical interpretation
as it supports a distinguishing (often termed ``separable'') set of two-valued states.
A ``canonical'' classical representation will be explicitly enumerated.

\item[(ii)] All classical interpretations of Hardy's configuration~(\ref{2020-hardy-e-hardyset}) enumerated in (i)
predict that, if the system is prepared in state $\Psi$, then the observable $dd$ never occurs.
That is, Hardy's setup is a gadget graph~\cite{tutte_1954,SZABO2009436,Ramanathan-18} with a ``true-implies-false (classical) set of two-valued states'' (TIFS).
Indeed, it is one out of three minimal nonisomorphic true-implies-false configurations in four dimensions~\cite[Fig.~4(a)]{2018-minimalYIYS}.
\end{enumerate}

Hardy's configuration~(\ref{2020-hardy-e-hardyset}) allows a classical interpretation because
it supports a set of 186 two-valued states that distinguishes  different observables from one another.
This means that the elements of every pair of distinct observables can be ``separated'' or ``distinguished'' by (at least)
one two-valued state such that the respective state values of these elements are different.
Therefore, by Kochen and Specker's Theorem~0~\cite{kochen1},
the structure of its observables can be embedded in some Boolean algebra, which indicates classical representability.

An explicit construction of a classical model of a propositional structure corresponding to Hardy's 1993 configuration~\cite{Hardy-93}
is enumerated in
Table~\ref{2020-hardy-tablepartitionlogic}. Its realization
is in terms of eight partitions (corresponding to the eight contexts) of the index set
$\{1,2,\ldots , 185,186\}$ of 186 two-valued states.
The elements of the partitions corresponding to the 21 atomic propositions
[which are obtained by ``completing'' the context as enumerated in
Eq.~(\ref{2020-hardy-e-hardyset})]
are the index sets of all two-valued states which obtain the value ``1'' on the respective atoms.
A detailed description of this construction can be found in Refs.~\cite{svozil-2001-eua,svozil-2018-b,svozil-2017-b}.
\begin{table*}
\begin{tabular}{ r l }
$\Psi$ =& $\{$1,2,3,4,5,6$\}$,
\\
2 =& $\{$7,8,9,10,11,12,13,14,15,16,17,18,19,20,21,22,23,24,25,26,27,28,29,30,31,32,33,34,35,36,37,38,39,\\&
40,41,42,43,44,45,46,47,48,49,50,51,52,53,54,55,56,57,58,59,60,61,62,63,64,65,66,67,68$\}$,
\\
3 =& $\{$69,70,71,72,73,74,75,76,77,78,79,80,81,82,83,84,85,86,87,88,89,90,\\&91,92,93,94,95,96,97,98,99,100,
101,102,103,104,105,106,107,108,109,110,111,112,\\&113,114,115,116,117,118,119,120,121,122,123,124,125,126,127,128,129,130$\}$,
\\
$vd$ =& $\{$131,132,133,134,135,136,137,138,139,140,141,142,143,144,145,146,147,148,149,150,\\&151,152,153,154,155,156,157,158,159,160,161,162,163,164,165,166,167,168,\\&169,170,171,172,173,174,175,176,177,178,179,180,181,182,183,184,185,186$\}$,
\\
$uu$ =& $\{$7,8,9,10,11,12,13,14,15,16,17,18,19,20,21,22,23,24,25,26,69,70,71,72,73,74,75,76,77,78,\\&79,80,81,82,83,84,85,86,87,88$\}$,
\\
$vu$ =& $\{$1,2,35,36,39,40,43,44,47,48,97,98,101,102,105,106,109,110,151,152,\\&153,154,155,160,161,162,163,164,169,170,171,172,173,178,179,180,181,182$\}$,
\\
$cv$ =& $\{$3,4,5,6,37,38,41,42,45,46,49,50,61,62,63,64,65,66,67,68,99,\\&100,103,104,107,108,111,112,123,124,125,126,127,128,129,130,156,157,158,159,\\&165,166,167,168,174,175,176,177,183,184,185,186$\}$,
\\
8 =& $\{$1,7,8,12,13,17,18,22,23,27,29,31,33,35,39,43,47,51,52,56,57,69,70,74,75,79,\\&80,84,85,89,91,93,95,97,101,105,109,113,114,118,119,131,132,136,\\&137,141,142,146,147,151,152,160,161,169,170,178,179$\}$,
\\
9 =& $\{$2,9,10,14,15,19,20,24,25,28,30,32,34,36,40,44,48,53,54,58,59,71,72,76,77,81,\\&82,86,87,90,92,94,96,98,102,106,110,115,116,120,121,133,\\&134,138,139,143,144,148,149,153,154,162,163,171,172,180,181$\}$,
\\
$dd$ =& $\{$11,16,21,26,55,60,73,78,83,88,117,122,135,140,145,150,155,164,173,182$\}$,
\\
11 =& $\{$5,7,9,12,14,17,19,22,24,51,53,56,58,61,63,65,67,69,71,74,76,79,81,84,86,113,115,\\&118,120,123,125,127,129,131,133,136,138,141,143,146,148,151,153,\\&156,158,160,162,165,167,169,171,174,176,178,180,183,185$\}$,
\\
12 =& $\{$6,8,10,13,15,18,20,23,25,52,54,57,59,62,64,66,68,70,72,75,77,80,82,85,87,\\&114,116,119,121,124,126,128,130,132,134,137,139,142,144,147,149,152,154,\\&157,159,161,163,166,168,170,172,175,177,179,181,184,186$\}$,
\\
$vc$ =& $\{$1,2,3,4,27,28,29,30,31,32,33,34,35,36,37,38,39,40,41,42,43,44,45,46,47,48,\\&49,50,89,90,91,92,93,94,95,96,97,98,99,100,101,102,103,104,105,106,\\&107,108,109,110,111,112$\}$,
\\
$uv$ =& $\{$5,6,51,52,53,54,55,56,57,58,59,60,61,62,63,64,65,66,67,68,113,114,115,116,\\&117,118,119,120,121,122,123,124,125,126,127,128,129,130$\}$,
\\
$dv$ =& $\{$27,28,29,30,31,32,33,34,51,52,53,54,55,56,57,58,59,60,89,90,91,92,93,94,95,96,\\&113,114,115,116,117,118,119,120,121,122,131,132,133,134,\\&135,136,137,138,139,140,141,142,143,144,145,146,147,148,149,150$\}$,
\\
16 =& $\{$7,8,9,10,11,12,13,14,15,16,35,36,37,38,39,40,41,42,61,62,63,64,69,70,71,\\&72,73,74,75,76,77,78,97,98,99,100,101,102,103,104,123,124,125,\\&126,151,152,153,154,155,156,157,158,159,160,161,162,163,164,165,166,167,168$\}$,
\\
17 =& $\{$17,18,19,20,21,22,23,24,25,26,43,44,45,46,47,48,49,50,65,66,67,68,79,80,81,\\&82,83,84,85,86,87,88,105,106,107,108,109,110,111,112,127,\\&128,129,130,169,170,171,172,173,174,175,176,177,178,179,180,181,182,183,184,185,186$\}$,
\\
18 =& $\{$3,7,8,9,10,11,17,18,19,20,21,27,28,31,32,37,41,45,49,69,70,71,72,73,79,\\&80,81,82,83,89,90,93,94,99,103,107,111,131,132,133,134,135,141,\\&142,143,144,145,156,157,165,166,174,175,183,184$\}$,
\\
19 =& $\{$4,12,13,14,15,16,22,23,24,25,26,29,30,33,34,38,42,46,50,74,75,76,77,78,\\&84,85,86,87,88,91,92,95,96,100,104,108,112,136,137,138,139,140,\\&146,147,148,149,150,158,159,167,168,176,177,185,186$\}$,
\\
20 =& $\{$27,28,29,30,35,36,37,38,43,44,45,46,51,52,53,54,55,61,62,65,66,89,90,91,\\&92,97,98,99,100,105,106,107,108,113,114,115,116,117,123,124,127,\\&128,131,132,133,134,135,136,137,138,139,140,151,152,153,154,155,\\&156,157,158,159,169,170,171,172,173,174,175,176,177$\}$,
\\
21 =& $\{$31,32,33,34,39,40,41,42,47,48,49,50,56,57,58,59,60,63,64,67,68,93,94,95,\\&96,101,102,103,104,109,110,111,112,118,119,120,121,122,125,126,\\&129,130,141,142,143,144,145,146,147,148,149,150,160,161,162,163,\\& 164,165,166,167,168,178,179,180,181,182,183,184,185,186$\}$.
\end{tabular}
\caption{\label{2020-hardy-tablepartitionlogic}
Partition logic representing classical probabilities of the Hardy configuration~\cite{Hardy-93},
whose intertwined contexts are enumerated in Eq.(~\ref{2020-hardy-e-hardyset}),
obtained from the separating or distinguishing set of all~186 two-valued states it supports.
Note that the intersection of
$\Psi \cap dd= \{1,2,3,4,5,6\} \cap \{11,16,21,26,55,60,73,78,83,88,117,122,135,140,145,150,155,164,173,182\}=\emptyset$
is empty, yielding true-implies-false relations among $\Psi$ and $dd$ and vice versa, respectively.}
\end{table*}

Next, we shall elaborate on a classical prediction which is violated by quantum predictions:
If $\Psi$ is assumed to be true---that is, if a classical system is prepared (also known as preselected)
in the state corresponding to observable $\Psi$---then the outcome corresponding to the observable $dd$ cannot occur.

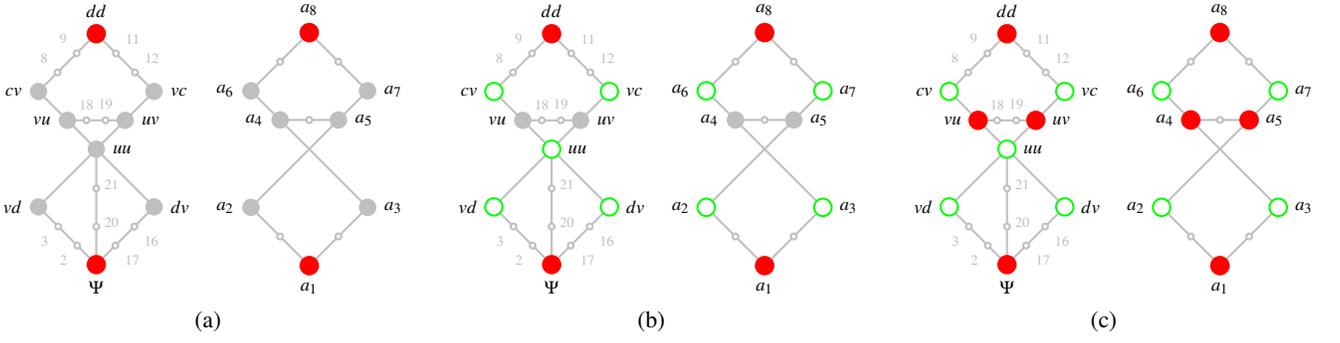
\begin{figure*}
\begin{center}
\begin{tabular}{ c c c c c c}
\resizebox{.15\textwidth}{!}{%
\begin{tikzpicture}  [scale=0.5]
\tikzstyle{every path}=[line width=1pt]
\newdimen\ms
\ms=0.1cm
\tikzstyle{s1}=[color=red,rectangle,inner sep=3.5]
\tikzstyle{c3}=[circle,inner sep={\ms/8},minimum size=4*\ms]
\tikzstyle{c2}=[circle,inner sep={\ms/8},minimum size=3*\ms]
\tikzstyle{c1}=[circle,inner sep={\ms/8},minimum size=2*\ms]
\tikzstyle{cs1}=[circle,inner sep={\ms/8},minimum size=1*\ms]
%
%
\coordinate (psi) at (2,0);
\coordinate (vd) at (0,2);
\coordinate (dv) at (4,2);
\coordinate (uu) at (2,4);
\coordinate (vu) at (1,5);
\coordinate (uv) at (3,5);
\coordinate (cv) at (0,6);
\coordinate (vc) at (4,6);
\coordinate (dd) at (2,8);
%
%
\draw [color=lightgray] (psi) -- (vd)  coordinate[cs1,fill=white,draw=lightgray,pos=0.33,label=below left:{\scriptsize \color{lightgray}2}] (2)  coordinate[cs1,fill=white,draw=lightgray,pos=0.66,label=below left:{\scriptsize \color{lightgray}3}] (3);
\draw [color=lightgray] (psi) -- (uu)  coordinate[cs1,fill=white,draw=lightgray,pos=0.33,label={[right,xshift=0.2mm]:{\scriptsize \color{lightgray}20}}] (20)  coordinate[cs1,fill=white,draw=lightgray,pos=0.66,label={[right,xshift=0.2mm]:{\scriptsize \color{lightgray}21}}] (21);
\draw [color=lightgray] (psi) -- (dv) coordinate[cs1,fill=white,draw=lightgray,pos=0.33,label=below right:{\scriptsize \color{lightgray}17}]  (17) coordinate[cs1,fill=white,draw=lightgray,pos=0.66,label=below right:{\scriptsize \color{lightgray}16}] (16);
\draw [color=lightgray] (vd) -- (vc);
\draw [color=lightgray] (dv) -- (cv);
\draw [color=lightgray] (vu) -- (uv) coordinate[cs1,fill=white,draw=lightgray,pos=0.33,label=above:{\scriptsize \color{lightgray}18}] (18)  coordinate[cs1,fill=white,draw=lightgray,pos=0.66,label=above:{\scriptsize \color{lightgray}19}] (19);
\draw [color=lightgray] (cv) -- (dd) coordinate[cs1,fill=white,draw=lightgray,pos=0.33,label=above left:{\scriptsize \color{lightgray}8}]  (8) coordinate[cs1,fill=white,draw=lightgray,pos=0.66,label=above left:{\scriptsize \color{lightgray}9}] (9);
\draw [color=lightgray] (vc) -- (dd) coordinate[cs1,fill=white,draw=lightgray,pos=0.33,label=above right:{\scriptsize \color{lightgray}12}] (12)  coordinate[cs1,fill=white,draw=lightgray,pos=0.66,label=above right:{\scriptsize \color{lightgray}11}] (12);
%
%
\draw (psi) coordinate[c2,fill=red,draw=red,label=below:$\Psi$];
\draw (vd) coordinate[c2,fill=lightgray,label=left:$vd$];
\draw (dv) coordinate[c2,fill=lightgray,label=right:$dv$];
\draw (uu) coordinate[c2,fill=lightgray,label=right:$uu$];
\draw (vu) coordinate[c2,fill=lightgray,label=left:$vu$];
\draw (uv) coordinate[c2,fill=lightgray,label=right:$uv$];
\draw (cv) coordinate[c2,fill=lightgray,label=left:$cv$];
\draw (vc) coordinate[c2,fill=lightgray,label=right:$vc$];
\draw (dd) coordinate[c2,fill=red,draw=red,label=above:$dd$];
\end{tikzpicture}
}
&
\resizebox{.15\textwidth}{!}{%
\begin{tikzpicture}  [scale=1]

\tikzstyle{every path}=[line width=1pt]

\newdimen\ms
\ms=0.1cm
\tikzstyle{s1}=[color=lightgray,rectangle,inner sep=3.5]
\tikzstyle{c3}=[circle,inner sep={\ms/8},minimum size=5*\ms]
\tikzstyle{c2}=[circle,inner sep={\ms/8},minimum size=3*\ms]
\tikzstyle{c1}=[circle,inner sep={\ms/8},minimum size=2*\ms]
\tikzstyle{cs1}=[circle,inner sep={\ms/8},minimum size=1*\ms]


\coordinate (a8) at  (1,2);
\coordinate (a7) at (2,1);
\coordinate (a4) at (0.5,0.5);
\coordinate (a3) at (2,-1);
\coordinate (a1) at (1,-2);
\coordinate (a2) at (0,-1);
\coordinate (a5) at (1.5,0.5);
\coordinate (a6) at (0,1);


\draw [color=lightgray] (a1) -- (a3)  coordinate[cs1,fill=white,draw=lightgray,pos=0.5] (2);
\draw [color=lightgray] (a3) -- (a6);
\draw [color=lightgray] (a6) -- (a8)  coordinate[cs1,fill=white,draw=lightgray,pos=0.5] (7);
\draw [color=lightgray] (a7) -- (a8)  coordinate[cs1,fill=white,draw=lightgray,pos=0.5] (3);
\draw [color=lightgray] (a2) -- (a7);
\draw [color=lightgray] (a2) -- (a1)  coordinate[cs1,fill=white,draw=lightgray,pos=0.5] (10);
\draw [color=lightgray] (a4) -- (a5)  coordinate[cs1,fill=white,draw=lightgray,pos=0.5] (13);


\draw (a1) coordinate[c2,fill=red,draw=red,label=below:$a_1$];
\draw (a2) coordinate[c2,fill=lightgray,label=left:$a_{2}$];
\draw (a3) coordinate[c2,fill=lightgray,label=right:$a_3$];
\draw (a4) coordinate[c2,fill=lightgray,label=left:$a_4$];
\draw (a5) coordinate[c2,fill=lightgray,label=right:$a_{5}$];
\draw (a6) coordinate[c2,fill=lightgray,label=left:$a_6$];
\draw (a7) coordinate[c2,fill=lightgray,label=right:$a_7$];
\draw (a8) coordinate[c2,fill=red,draw=red,label=above:$a_8$];

\end{tikzpicture}
}
\quad
&
\quad
\resizebox{.15\textwidth}{!}{%
\begin{tikzpicture}  [scale=0.5]

\tikzstyle{every path}=[line width=1pt]

\newdimen\ms
\ms=0.1cm
\tikzstyle{s1}=[color=red,rectangle,inner sep=3.5]
\tikzstyle{c3}=[circle,inner sep={\ms/8},minimum size=4*\ms]
\tikzstyle{c2}=[circle,inner sep={\ms/8},minimum size=3*\ms]
\tikzstyle{c1}=[circle,inner sep={\ms/8},minimum size=2*\ms]
\tikzstyle{cs1}=[circle,inner sep={\ms/8},minimum size=1*\ms]


\coordinate (psi) at (2,0);
\coordinate (vd) at (0,2);
\coordinate (dv) at (4,2);
\coordinate (uu) at (2,4);
\coordinate (vu) at (1,5);
\coordinate (uv) at (3,5);
\coordinate (cv) at (0,6);
\coordinate (vc) at (4,6);
\coordinate (dd) at (2,8);


\draw [color=lightgray] (psi) -- (vd)  coordinate[cs1,fill=white,draw=lightgray,pos=0.33,label=below left:{\scriptsize \color{lightgray}2}] (2)  coordinate[cs1,fill=white,draw=lightgray,pos=0.66,label=below left:{\scriptsize \color{lightgray}3}] (3);
\draw [color=lightgray] (psi) -- (uu)  coordinate[cs1,fill=white,draw=lightgray,pos=0.33,label={[right,xshift=0.2mm]:{\scriptsize \color{lightgray}20}}] (20)  coordinate[cs1,fill=white,draw=lightgray,pos=0.66,label={[right,xshift=0.2mm]:{\scriptsize \color{lightgray}21}}] (21);
\draw [color=lightgray] (psi) -- (dv) coordinate[cs1,fill=white,draw=lightgray,pos=0.33,label=below right:{\scriptsize \color{lightgray}17}]  (17) coordinate[cs1,fill=white,draw=lightgray,pos=0.66,label=below right:{\scriptsize \color{lightgray}16}] (16);
\draw [color=lightgray] (vd) -- (vc);
\draw [color=lightgray] (dv) -- (cv);
\draw [color=lightgray] (vu) -- (uv) coordinate[cs1,fill=white,draw=lightgray,pos=0.33,label=above:{\scriptsize \color{lightgray}18}] (18)  coordinate[cs1,fill=white,draw=lightgray,pos=0.66,label=above:{\scriptsize \color{lightgray}19}] (19);
\draw [color=lightgray] (cv) -- (dd) coordinate[cs1,fill=white,draw=lightgray,pos=0.33,label=above left:{\scriptsize \color{lightgray}8}]  (8) coordinate[cs1,fill=white,draw=lightgray,pos=0.66,label=above left:{\scriptsize \color{lightgray}9}] (9);
\draw [color=lightgray] (vc) -- (dd) coordinate[cs1,fill=white,draw=lightgray,pos=0.33,label=above right:{\scriptsize \color{lightgray}12}] (12)  coordinate[cs1,fill=white,draw=lightgray,pos=0.66,label=above right:{\scriptsize \color{lightgray}11}] (12);


\draw (psi) coordinate[c2,fill=red,draw=red,label=below:$\Psi$];
\draw (vd) coordinate[c2,fill=white,draw=green,label=left:$vd$];
\draw (dv) coordinate[c2,fill=white,draw=green,label=right:$dv$];
\draw (uu) coordinate[c2,fill=white,draw=green,label=right:$uu$];
\draw (vu) coordinate[c2,fill=lightgray,label=left:$vu$];
\draw (uv) coordinate[c2,fill=lightgray,label=right:$uv$];
\draw (cv) coordinate[c2,fill=white,draw=green,label=left:$cv$];
\draw (vc) coordinate[c2,fill=white,draw=green,label=right:$vc$];
\draw (dd) coordinate[c2,fill=red,draw=red,label=above:$dd$];

\end{tikzpicture}
}
&
\resizebox{.15\textwidth}{!}{%
\begin{tikzpicture}  [scale=1]
\tikzstyle{every path}=[line width=1pt]

\newdimen\ms
\ms=0.1cm
\tikzstyle{s1}=[color=lightgray,rectangle,inner sep=3.5]
\tikzstyle{c3}=[circle,inner sep={\ms/8},minimum size=5*\ms]
\tikzstyle{c2}=[circle,inner sep={\ms/8},minimum size=3*\ms]
\tikzstyle{c1}=[circle,inner sep={\ms/8},minimum size=2*\ms]
\tikzstyle{cs1}=[circle,inner sep={\ms/8},minimum size=1*\ms]


\coordinate (a8) at  (1,2);
\coordinate (a7) at (2,1);
\coordinate (a4) at (0.5,0.5);
\coordinate (a3) at (2,-1);
\coordinate (a1) at (1,-2);
\coordinate (a2) at (0,-1);
\coordinate (a5) at (1.5,0.5);
\coordinate (a6) at (0,1);


\draw [color=lightgray] (a1) -- (a3)  coordinate[cs1,fill=white,draw=lightgray,pos=0.5] (2);
\draw [color=lightgray] (a3) -- (a6);
\draw [color=lightgray] (a6) -- (a8)  coordinate[cs1,fill=white,draw=lightgray,pos=0.5] (7);
\draw [color=lightgray] (a7) -- (a8)  coordinate[cs1,fill=white,draw=lightgray,pos=0.5] (3);
\draw [color=lightgray] (a2) -- (a7);
\draw [color=lightgray] (a2) -- (a1)  coordinate[cs1,fill=white,draw=lightgray,pos=0.5] (10);
\draw [color=lightgray] (a4) -- (a5)  coordinate[cs1,fill=white,draw=lightgray,pos=0.5] (13);


\draw (a1) coordinate[c2,fill=red,draw=red,label=below:$a_1$];
\draw (a2) coordinate[c2,fill=white,draw=green,label=left:$a_{2}$];
\draw (a3) coordinate[c2,fill=white,draw=green,label=right:$a_3$];
\draw (a4) coordinate[c2,fill=lightgray,label=left:$a_4$];
\draw (a5) coordinate[c2,fill=lightgray,label=right:$a_{5}$];
\draw (a6) coordinate[c2,fill=white,draw=green,label=left:$a_6$];
\draw (a7) coordinate[c2,fill=white,draw=green,label=right:$a_7$];
\draw (a8) coordinate[c2,fill=red,draw=red,label=above:$a_8$];

\end{tikzpicture}
}
\quad
&
\quad
\resizebox{.15\textwidth}{!}{%
\begin{tikzpicture}  [scale=0.5]

\tikzstyle{every path}=[line width=1pt]

\newdimen\ms
\ms=0.1cm
\tikzstyle{s1}=[color=red,rectangle,inner sep=3.5]
\tikzstyle{c3}=[circle,inner sep={\ms/8},minimum size=4*\ms]
\tikzstyle{c2}=[circle,inner sep={\ms/8},minimum size=3*\ms]
\tikzstyle{c1}=[circle,inner sep={\ms/8},minimum size=2*\ms]
\tikzstyle{cs1}=[circle,inner sep={\ms/8},minimum size=1*\ms]


\coordinate (psi) at (2,0);
\coordinate (vd) at (0,2);
\coordinate (dv) at (4,2);
\coordinate (uu) at (2,4);
\coordinate (vu) at (1,5);
\coordinate (uv) at (3,5);
\coordinate (cv) at (0,6);
\coordinate (vc) at (4,6);
\coordinate (dd) at (2,8);


\draw [color=lightgray] (psi) -- (vd)  coordinate[cs1,fill=white,draw=lightgray,pos=0.33,label=below left:{\scriptsize \color{lightgray}2}] (2)  coordinate[cs1,fill=white,draw=lightgray,pos=0.66,label=below left:{\scriptsize \color{lightgray}3}] (3);
\draw [color=lightgray] (psi) -- (uu)  coordinate[cs1,fill=white,draw=lightgray,pos=0.33,label={[right,xshift=0.2mm]:{\scriptsize \color{lightgray}20}}] (20)  coordinate[cs1,fill=white,draw=lightgray,pos=0.66,label={[right,xshift=0.2mm]:{\scriptsize \color{lightgray}21}}] (21);
\draw [color=lightgray] (psi) -- (dv) coordinate[cs1,fill=white,draw=lightgray,pos=0.33,label=below right:{\scriptsize \color{lightgray}17}]  (17) coordinate[cs1,fill=white,draw=lightgray,pos=0.66,label=below right:{\scriptsize \color{lightgray}16}] (16);
\draw [color=lightgray] (vd) -- (vc);
\draw [color=lightgray] (dv) -- (cv);
\draw [color=lightgray] (vu) -- (uv) coordinate[cs1,fill=white,draw=lightgray,pos=0.33,label=above:{\scriptsize \color{lightgray}18}] (18)  coordinate[cs1,fill=white,draw=lightgray,pos=0.66,label=above:{\scriptsize \color{lightgray}19}] (19);
\draw [color=lightgray] (cv) -- (dd) coordinate[cs1,fill=white,draw=lightgray,pos=0.33,label=above left:{\scriptsize \color{lightgray}8}]  (8) coordinate[cs1,fill=white,draw=lightgray,pos=0.66,label=above left:{\scriptsize \color{lightgray}9}] (9);
\draw [color=lightgray] (vc) -- (dd) coordinate[cs1,fill=white,draw=lightgray,pos=0.33,label=above right:{\scriptsize \color{lightgray}12}] (12)  coordinate[cs1,fill=white,draw=lightgray,pos=0.66,label=above right:{\scriptsize \color{lightgray}11}] (12);


\draw (psi) coordinate[c2,fill=red,draw=red,label=below:$\Psi$];
\draw (vd) coordinate[c2,fill=white,draw=green,label=left:$vd$];
\draw (dv) coordinate[c2,fill=white,draw=green,label=right:$dv$];
\draw (uu) coordinate[c2,fill=white,draw=green,label=right:$uu$];
\draw (vu) coordinate[c2,fill=red,draw=red,label=left:$vu$];
\draw (uv) coordinate[c2,fill=red,draw=red,label=right:$uv$];
\draw (cv) coordinate[c2,fill=white,draw=green,label=left:$cv$];
\draw (vc) coordinate[c2,fill=white,draw=green,label=right:$vc$];
\draw (dd) coordinate[c2,fill=red,draw=red,label=above:$dd$];

\end{tikzpicture}
}
&
\resizebox{.15\textwidth}{!}{%
\begin{tikzpicture}  [scale=1]
\tikzstyle{every path}=[line width=1pt]

\newdimen\ms
\ms=0.1cm
\tikzstyle{s1}=[color=lightgray,rectangle,inner sep=3.5]
\tikzstyle{c3}=[circle,inner sep={\ms/8},minimum size=5*\ms]
\tikzstyle{c2}=[circle,inner sep={\ms/8},minimum size=3*\ms]
\tikzstyle{c1}=[circle,inner sep={\ms/8},minimum size=2*\ms]
\tikzstyle{cs1}=[circle,inner sep={\ms/8},minimum size=1*\ms]


\coordinate (a8) at  (1,2);
\coordinate (a7) at (2,1);
\coordinate (a4) at (0.5,0.5);
\coordinate (a3) at (2,-1);
\coordinate (a1) at (1,-2);
\coordinate (a2) at (0,-1);
\coordinate (a5) at (1.5,0.5);
\coordinate (a6) at (0,1);


\draw [color=lightgray] (a1) -- (a3)  coordinate[cs1,fill=white,draw=lightgray,pos=0.5] (2);
\draw [color=lightgray] (a3) -- (a6);
\draw [color=lightgray] (a6) -- (a8)  coordinate[cs1,fill=white,draw=lightgray,pos=0.5] (7);
\draw [color=lightgray] (a7) -- (a8)  coordinate[cs1,fill=white,draw=lightgray,pos=0.5] (3);
\draw [color=lightgray] (a2) -- (a7);
\draw [color=lightgray] (a2) -- (a1)  coordinate[cs1,fill=white,draw=lightgray,pos=0.5] (10);
\draw [color=lightgray] (a4) -- (a5)  coordinate[cs1,fill=white,draw=lightgray,pos=0.5] (13);


\draw (a1) coordinate[c2,fill=red,draw=red,label=below:$a_1$];
\draw (a2) coordinate[c2,fill=white,draw=green,label=left:$a_{2}$];
\draw (a3) coordinate[c2,fill=white,draw=green,label=right:$a_3$];
\draw (a4) coordinate[c2,fill=red,draw=red,label=left:$a_4$];
\draw (a5) coordinate[c2,fill=red,draw=red,label=right:$a_{5}$];
\draw (a6) coordinate[c2,fill=white,draw=green,label=left:$a_6$];
\draw (a7) coordinate[c2,fill=white,draw=green,label=right:$a_7$];
\draw (a8) coordinate[c2,fill=red,draw=red,label=above:$a_8$];

\end{tikzpicture}
}
\\
\multicolumn{2}{c}{(a)}&\multicolumn{2}{c}{(b)}&\multicolumn{2}{c}{(c)}
\end{tabular}
\end{center}
\caption{\label{2020-hardy-fig1proof}
Graphical presentation of a three-step  proof by contradiction that from the pairs of observables $\{\Psi,dd\}$ and $\{a_1,a_8\}$, only one element can have assigned the classical value 1:
(a) suppose otherwise; that is,  $\Psi= dd=1$ and $a_1=a_8=1$;
(b) then, by exclusivity,  $vd= dv=uu=vc=cv=0$ and $a_2=a_3=a_6=a_7=0$;
(c) then, by completeness,  $vu=uv=1$ and $a_4=a5=1$, contradicting exclusivity.
Small circles indicate ``auxiliary'' observables which can be chosen freely,
subject to orthogonality constraints: all smooth lines indicate respective contexts representing orthonormal bases.
}
\end{figure*}

For a proof by contradiction depicted in Fig.~\ref{2020-hardy-fig1proof}
(wrongly) suppose that both $\Psi$ as well as $dd$ were both true simultaneously.
Then by the standard admissibility criteria for two-valued states~\cite{Greechie1974,2015-AnalyticKS}
(also denoted as completeness and exclusivity~\cite{cabello-2013-beg,Cabello-2014-gtatqc,Xu-Chen-Guehnw-2020}),
$cv=vc=vd=dv=uu=0$, enforcing $vu=uv=1$ which contradicts admissibility (completeness and exclusivity).

The only remaining possibility is that $\psi$ and $dd$ have opposite values if one of them is true (they still may both be 0).
Therefore, any two-valued state for which $\Psi$ is 1 -- that is, in which the observable corresponding to
$\Psi$ occurs -- must classically result in nonoccurrence of the outcome corresponding to the observable $dd$; and vice versa.
This particular relation between the input and output ports of gadget graphs~\cite{svozil-2020-c} has been called
1-0--property~\cite{svozil-2006-omni}, or one dominated by a
true-implies-false set of two-valued states (TIFS)~\cite{2018-minimalYIYS}.

As mentioned earlier, the first true-implies-false gadget seems
to have been introduced by Kochen and Specker~\cite[Fig.~1, p.~182]{kochen2}
and used by them as a subgraph of~$\Gamma_1$~\cite[p.~68]{kochen1}
in three dimensions.
Its orthogonality hypergraph is depicted in Fig.~\ref{2020-hardy-fig1}(b).
As mentioned earlier this gadget seems to have been independently discussed by, among others,
Pitowsky, who called it the ``cat's cradle''~\cite{Pitowsky2003395,pitowsky-06}.
See also
Fig.~1 in~\cite[p.~123]{Greechie1974} (reprinted in Ref.~\cite{Greechie-Suppes1976}),
a subgraph in Fig.~21   in Ref.~\cite[pp.~126-127]{redhead},
Fig.~B.l   in \cite[p.~64]{Belinfante-73},
\cite[pp.~588-589]{stairs83},
Fig.~2  in \cite[p.~446]{clifton-93},  and
Fig.~2.4.6 in \cite[p.~39]{pulmannova-91} for early discussions of the true-implies-false prediction.

\begin{figure}[htb]
\begin{center}
\begin{tabular}{ c c c c c c }
%
%
\resizebox{.07\textwidth}{!}{%
\begin{tikzpicture}  [scale=0.5]
\tikzstyle{every path}=[line width=1pt]
\newdimen\ms
\ms=0.1cm
\tikzstyle{c2}=[circle,inner sep={\ms/8},minimum size=2*\ms]


\coordinate (psi) at (2,0);
\coordinate (vd) at (0,2);
\coordinate (dv) at (4,2);
\coordinate (uu) at (2,4);
\coordinate (vu) at (1,5);
\coordinate (uv) at (3,5);
\coordinate (cv) at (0,6);
\coordinate (vc) at (4,6);
\coordinate (dd) at (2,8);

\node[draw] at (2,6.25) {\Large \bf 1};


\draw [color=lightgray] (psi) -- (vd)  coordinate[c2,fill=white,draw=green,pos=0.33] (2)  coordinate[c2,fill=white,draw=green,pos=0.66] (3);
\draw [color=lightgray] (psi) -- (uu)  coordinate[c2,fill=white,draw=green,pos=0.33] (20)  coordinate[c2,fill=white,draw=green,pos=0.66] (21);
\draw [color=lightgray] (psi) -- (dv)  coordinate[c2,fill=white,draw=green,pos=0.33]  (17) coordinate[c2,fill=white,draw=green,pos=0.66] (16);
\draw [color=lightgray] (vd) -- (vc);
\draw [color=lightgray] (dv) -- (cv);
\draw [color=lightgray] (vu) -- (uv)  coordinate[c2,fill=white,draw=green,pos=0.33] (18)  coordinate[c2,fill=white,draw=green,pos=0.66] (19);
\draw [color=lightgray] (cv) -- (dd)  coordinate[c2,fill=red,draw=red,pos=0.33]  (8) coordinate[c2,fill=white,draw=green,pos=0.66] (9);
\draw [color=olive] (vc) -- (dd)  coordinate[c2,fill=white,draw=green,pos=0.33] (11)  coordinate[c2,fill=white,draw=green,pos=0.66] (12);


\draw (psi) coordinate[c2,fill=red,draw=red,label=below:{\Large $\Psi$}];
\draw (vd) coordinate[c2,fill=white,draw=green];
\draw (dv) coordinate[c2,fill=white,draw=green];
\draw (uu) coordinate[c2,fill=white,draw=green];
\draw (vu) coordinate[c2,fill=red,draw=red];
\draw (uv) coordinate[c2,fill=white,draw=green];
\draw (cv) coordinate[c2,fill=white,draw=green];
\draw (vc) coordinate[c2,fill=red,draw=red];
\draw (dd) coordinate[c2,fill=white,draw=green,label=above:{\Large $dd$}];

\end{tikzpicture}
}
\hfill & \hfill
%
%
\resizebox{.07\textwidth}{!}{%
\begin{tikzpicture}  [scale=0.5]
\tikzstyle{every path}=[line width=1pt]
\newdimen\ms
\ms=0.1cm
\tikzstyle{c2}=[circle,inner sep={\ms/8},minimum size=2*\ms]


\coordinate (psi) at (2,0);
\coordinate (vd) at (0,2);
\coordinate (dv) at (4,2);
\coordinate (uu) at (2,4);
\coordinate (vu) at (1,5);
\coordinate (uv) at (3,5);
\coordinate (cv) at (0,6);
\coordinate (vc) at (4,6);
\coordinate (dd) at (2,8);

\node[draw] at (2,6.25) {\Large \bf 2};


\draw [color=lightgray] (psi) -- (vd)  coordinate[c2,fill=white,draw=green,pos=0.33] (2)  coordinate[c2,fill=white,draw=green,pos=0.66] (3);
\draw [color=lightgray] (psi) -- (uu)  coordinate[c2,fill=white,draw=green,pos=0.33] (20)  coordinate[c2,fill=white,draw=green,pos=0.66] (21);
\draw [color=lightgray] (psi) -- (dv)  coordinate[c2,fill=white,draw=green,pos=0.33]  (17) coordinate[c2,fill=white,draw=green,pos=0.66] (16);
\draw [color=lightgray] (vd) -- (vc);
\draw [color=lightgray] (dv) -- (cv);
\draw [color=lightgray] (vu) -- (uv)  coordinate[c2,fill=white,draw=green,pos=0.33] (18)  coordinate[c2,fill=white,draw=green,pos=0.66] (19);
\draw [color=lightgray] (cv) -- (dd)  coordinate[c2,fill=white,draw=green,pos=0.33]  (8) coordinate[c2,fill=red,draw=red,pos=0.66] (9);
\draw [color=olive] (vc) -- (dd)  coordinate[c2,fill=white,draw=green,pos=0.33] (11)  coordinate[c2,fill=white,draw=green,pos=0.66] (12);


\draw (psi) coordinate[c2,fill=red,draw=red,label=below:{\Large $\Psi$}];
\draw (vd) coordinate[c2,fill=white,draw=green];
\draw (dv) coordinate[c2,fill=white,draw=green];
\draw (uu) coordinate[c2,fill=white,draw=green];
\draw (vu) coordinate[c2,fill=red,draw=red];
\draw (uv) coordinate[c2,fill=white,draw=green];
\draw (cv) coordinate[c2,fill=white,draw=green];
\draw (vc) coordinate[c2,fill=red,draw=red];
\draw (dd) coordinate[c2,fill=white,draw=green,label=above:{\Large $dd$}];

\end{tikzpicture}
}
\hfill & \hfill
%
%
\resizebox{.07\textwidth}{!}{%
\begin{tikzpicture}  [scale=0.5]
\tikzstyle{every path}=[line width=1pt]
\newdimen\ms
\ms=0.1cm
\tikzstyle{c2}=[circle,inner sep={\ms/8},minimum size=2*\ms]


\coordinate (psi) at (2,0);
\coordinate (vd) at (0,2);
\coordinate (dv) at (4,2);
\coordinate (uu) at (2,4);
\coordinate (vu) at (1,5);
\coordinate (uv) at (3,5);
\coordinate (cv) at (0,6);
\coordinate (vc) at (4,6);
\coordinate (dd) at (2,8);

\node[draw] at (2,6.25) {\Large \bf 3};


\draw [color=lightgray] (psi) -- (vd)  coordinate[c2,fill=white,draw=green,pos=0.33] (2)  coordinate[c2,fill=white,draw=green,pos=0.66] (3);
\draw [color=lightgray] (psi) -- (uu)  coordinate[c2,fill=white,draw=green,pos=0.33] (20)  coordinate[c2,fill=white,draw=green,pos=0.66] (21);
\draw [color=lightgray] (psi) -- (dv)  coordinate[c2,fill=white,draw=green,pos=0.33]  (17) coordinate[c2,fill=white,draw=green,pos=0.66] (16);
\draw [color=lightgray] (vd) -- (vc);
\draw [color=lightgray] (dv) -- (cv);
\draw [color=lightgray] (vu) -- (uv)  coordinate[c2,fill=red,draw=red,pos=0.33] (18)  coordinate[c2,fill=white,draw=green,pos=0.66] (19);
\draw [color=lightgray] (cv) -- (dd)  coordinate[c2,fill=white,draw=green,pos=0.33]  (8) coordinate[c2,fill=white,draw=green,pos=0.66] (9);
\draw [color=olive] (vc) -- (dd)  coordinate[c2,fill=white,draw=green,pos=0.33] (11)  coordinate[c2,fill=white,draw=green,pos=0.66] (12);


\draw (psi) coordinate[c2,fill=red,draw=red,label=below:{\Large $\Psi$}];
\draw (vd) coordinate[c2,fill=white,draw=green];
\draw (dv) coordinate[c2,fill=white,draw=green];
\draw (uu) coordinate[c2,fill=white,draw=green];
\draw (vu) coordinate[c2,fill=white,draw=green];
\draw (uv) coordinate[c2,fill=white,draw=green];
\draw (cv) coordinate[c2,fill=red,draw=red];
\draw (vc) coordinate[c2,fill=red,draw=red];
\draw (dd) coordinate[c2,fill=white,draw=green,label=above:{\Large $dd$}];

\end{tikzpicture}
}
&
%
%
\resizebox{.07\textwidth}{!}{%
\begin{tikzpicture}  [scale=0.5]
\tikzstyle{every path}=[line width=1pt]
\newdimen\ms
\ms=0.1cm
\tikzstyle{c2}=[circle,inner sep={\ms/8},minimum size=2*\ms]


\coordinate (psi) at (2,0);
\coordinate (vd) at (0,2);
\coordinate (dv) at (4,2);
\coordinate (uu) at (2,4);
\coordinate (vu) at (1,5);
\coordinate (uv) at (3,5);
\coordinate (cv) at (0,6);
\coordinate (vc) at (4,6);
\coordinate (dd) at (2,8);

\node[draw] at (2,6.25) {\Large \bf 4};


\draw [color=lightgray] (psi) -- (vd)  coordinate[c2,fill=white,draw=green,pos=0.33] (2)  coordinate[c2,fill=white,draw=green,pos=0.66] (3);
\draw [color=lightgray] (psi) -- (uu)  coordinate[c2,fill=white,draw=green,pos=0.33] (20)  coordinate[c2,fill=white,draw=green,pos=0.66] (21);
\draw [color=lightgray] (psi) -- (dv)  coordinate[c2,fill=white,draw=green,pos=0.33]  (17) coordinate[c2,fill=white,draw=green,pos=0.66] (16);
\draw [color=lightgray] (vd) -- (vc);
\draw [color=lightgray] (dv) -- (cv);
\draw [color=lightgray] (vu) -- (uv)  coordinate[c2,fill=white,draw=green,pos=0.33] (18)  coordinate[c2,fill=red,draw=red,pos=0.66] (19);
\draw [color=lightgray] (cv) -- (dd)  coordinate[c2,fill=white,draw=green,pos=0.33]  (8) coordinate[c2,fill=white,draw=green,pos=0.66] (9);
\draw [color=olive] (vc) -- (dd)  coordinate[c2,fill=white,draw=green,pos=0.33] (11)  coordinate[c2,fill=white,draw=green,pos=0.66] (12);


\draw (psi) coordinate[c2,fill=red,draw=red,label=below:{\Large $\Psi$}];
\draw (vd) coordinate[c2,fill=white,draw=green];
\draw (dv) coordinate[c2,fill=white,draw=green];
\draw (uu) coordinate[c2,fill=white,draw=green];
\draw (vu) coordinate[c2,fill=white,draw=green];
\draw (uv) coordinate[c2,fill=white,draw=green];
\draw (cv) coordinate[c2,fill=red,draw=red];
\draw (vc) coordinate[c2,fill=red,draw=red];
\draw (dd) coordinate[c2,fill=white,draw=green,label=above:{\Large $dd$}];

\end{tikzpicture}
}
\hfill & \hfill
%
%
\resizebox{.07\textwidth}{!}{%
\begin{tikzpicture}  [scale=0.5]
\tikzstyle{every path}=[line width=1pt]
\newdimen\ms
\ms=0.1cm
\tikzstyle{c2}=[circle,inner sep={\ms/8},minimum size=2*\ms]


\coordinate (psi) at (2,0);
\coordinate (vd) at (0,2);
\coordinate (dv) at (4,2);
\coordinate (uu) at (2,4);
\coordinate (vu) at (1,5);
\coordinate (uv) at (3,5);
\coordinate (cv) at (0,6);
\coordinate (vc) at (4,6);
\coordinate (dd) at (2,8);

\node[draw] at (2,6.25) {\Large \bf 5};


\draw [color=lightgray] (psi) -- (vd)  coordinate[c2,fill=white,draw=green,pos=0.33] (2)  coordinate[c2,fill=white,draw=green,pos=0.66] (3);
\draw [color=lightgray] (psi) -- (uu)  coordinate[c2,fill=white,draw=green,pos=0.33] (20)  coordinate[c2,fill=white,draw=green,pos=0.66] (21);
\draw [color=lightgray] (psi) -- (dv)  coordinate[c2,fill=white,draw=green,pos=0.33]  (17) coordinate[c2,fill=white,draw=green,pos=0.66] (16);
\draw [color=lightgray] (vd) -- (vc);
\draw [color=lightgray] (dv) -- (cv);
\draw [color=lightgray] (vu) -- (uv)  coordinate[c2,fill=white,draw=green,pos=0.33] (18)  coordinate[c2,fill=white,draw=green,pos=0.66] (19);
\draw [color=lightgray] (cv) -- (dd)  coordinate[c2,fill=white,draw=green,pos=0.33]  (8) coordinate[c2,fill=white,draw=green,pos=0.66] (9);
\draw [color=olive] (vc) -- (dd)  coordinate[c2,fill=red,draw=red,pos=0.33] (11)  coordinate[c2,fill=white,draw=green,pos=0.66] (12);


\draw (psi) coordinate[c2,fill=red,draw=red,label=below:{\Large $\Psi$}];
\draw (vd) coordinate[c2,fill=white,draw=green];
\draw (dv) coordinate[c2,fill=white,draw=green];
\draw (uu) coordinate[c2,fill=white,draw=green];
\draw (vu) coordinate[c2,fill=white,draw=green];
\draw (uv) coordinate[c2,fill=red,draw=red];
\draw (cv) coordinate[c2,fill=red,draw=red];
\draw (vc) coordinate[c2,fill=white,draw=green];
\draw (dd) coordinate[c2,fill=white,draw=green,label=above:{\Large $dd$}];

\end{tikzpicture}
}
\hfill & \hfill
%
%
\resizebox{.07\textwidth}{!}{%
\begin{tikzpicture}  [scale=0.5]
\tikzstyle{every path}=[line width=1pt]
\newdimen\ms
\ms=0.1cm
\tikzstyle{c2}=[circle,inner sep={\ms/8},minimum size=2*\ms]


\coordinate (psi) at (2,0);
\coordinate (vd) at (0,2);
\coordinate (dv) at (4,2);
\coordinate (uu) at (2,4);
\coordinate (vu) at (1,5);
\coordinate (uv) at (3,5);
\coordinate (cv) at (0,6);
\coordinate (vc) at (4,6);
\coordinate (dd) at (2,8);

\node[draw] at (2,6.25) {\Large \bf 6};


\draw [color=lightgray] (psi) -- (vd)  coordinate[c2,fill=white,draw=green,pos=0.33] (2)  coordinate[c2,fill=white,draw=green,pos=0.66] (3);
\draw [color=lightgray] (psi) -- (uu)  coordinate[c2,fill=white,draw=green,pos=0.33] (20)  coordinate[c2,fill=white,draw=green,pos=0.66] (21);
\draw [color=lightgray] (psi) -- (dv)  coordinate[c2,fill=white,draw=green,pos=0.33]  (17) coordinate[c2,fill=white,draw=green,pos=0.66] (16);
\draw [color=lightgray] (vd) -- (vc);
\draw [color=lightgray] (dv) -- (cv);
\draw [color=lightgray] (vu) -- (uv)  coordinate[c2,fill=white,draw=green,pos=0.33] (18)  coordinate[c2,fill=white,draw=green,pos=0.66] (19);
\draw [color=lightgray] (cv) -- (dd)  coordinate[c2,fill=white,draw=green,pos=0.33]  (8) coordinate[c2,fill=white,draw=green,pos=0.66] (9);
\draw [color=olive] (vc) -- (dd)  coordinate[c2,fill=white,draw=green,pos=0.33] (11)  coordinate[c2,fill=red,draw=red,pos=0.66] (12);


\draw (psi) coordinate[c2,fill=red,draw=red,label=below:{\Large $\Psi$}];
\draw (vd) coordinate[c2,fill=white,draw=green];
\draw (dv) coordinate[c2,fill=white,draw=green];
\draw (uu) coordinate[c2,fill=white,draw=green];
\draw (vu) coordinate[c2,fill=white,draw=green];
\draw (uv) coordinate[c2,fill=red,draw=red];
\draw (cv) coordinate[c2,fill=red,draw=red];
\draw (vc) coordinate[c2,fill=white,draw=green];
\draw (dd) coordinate[c2,fill=white,draw=green,label=above:{\Large $dd$}];

\end{tikzpicture}
}
\\
\end{tabular}
\end{center}
\caption{\label{2020-hardy-fig2}
Orthogonality hypergraphs of the  Hardy gadget with overlaid six two-valued states which are 1 at $\Psi$.}
\end{figure}
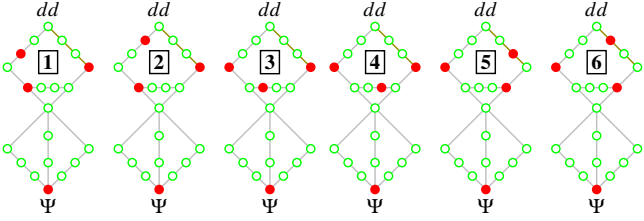

The full nuances of the predictions are revealed when the classical probabilities are computed.
As the classical probability distributions are just the convex combinations of all two-valued states~\cite[Chapter~2]{pitowsky},
it is easy to read them off from the
canonical partition logic enumerated in Table~\ref{2020-hardy-tablepartitionlogic}.
In particular, the true-implies-false gadget behavior at the terminals $\Psi$ and $dd$ can be directly read off from
\begin{equation}
\begin{split}
P_\Psi = \sum_{i \in \Psi} \lambda_i =  \lambda_1+\lambda_2+ \cdots + \lambda_6, \textrm{ and }\\
P_{dd} = \sum_{i \in dd} \lambda_i =  \lambda_{11}+\lambda_{16}+ \cdots + \lambda_{182}, \\
\textrm{with }\lambda_i\ge 0,\textrm{ and }\sum_{i=1}^{186} \lambda_i =1.
\end{split}
\end{equation}
Since the intersection of the index sets $\Psi$ and $dd$ is empty, $P_{dd}=0$ whenever $P_\Psi=1$, and vice versa.
For the sake of the example all six two-valued measures assigning 1 to $\Psi$ are depicted in Fig.~\ref{2020-hardy-fig2}.

One equivalent alternative way to characterize the classical probabilities completely would be to
exploit the Minkowski-Weyl ``main'' representation
theorem~\cite{ziegler,Henk-Ziegler-polytopes,Avis:1997:GCH:280651.280652,mcmullen-71,Schrijver,gruenbaum-2003,Fukuda-techrep}
and consider the classical convex polytope spanned by the 186  21-dimensional vectors whose components
are the values in $\{0,1\}$  of the two-valued states on the atomic propositions of the Hardy gadget.
From these vertices  (V-representation), the 35 half-spaces that are the bounds of the polytope
(H-representation) can be computed~\cite{froissart-81,pitowsky}. But due to space restrictions we omit this discussion,
although it might reveal quantum violations of Boole's (classical) ``conditions of experience''~\cite{Pit-94}.

\subsection{Original quantum realization}

Hardy's original quantum realization in terms of a particular type of faithful orthogonal representation
is quite involved, but for the sake of delineating it
we shall mostly stick to the nomenclature of the 1993 paper~\cite{Hardy-93}.
Suppose two two-state particles, and, for each one of the two particles,
consider three orthonormal bases of its two-dimensional Hilbert space, namely
\begin{equation}
\begin{aligned}
B_1 &= \{ \vert + \rangle , \vert - \rangle \} \equiv \{ {\bf e}_1 , {\bf e}_2 \}  ,             \\
B_2 &= \{ \vert u \rangle , \vert v \rangle \} \equiv \{ {\bf f}_1 , {\bf f}_2 \}  ,\textrm{ and}\\
B_3 &= \{ \vert c \rangle , \vert d \rangle \} \equiv \{ {\bf g}_1 , {\bf g}_2 \}  .
\end{aligned}
\label{2020-hardy-eq-b}
\end{equation}
The components of the respective unitary transformations ``rotating'' these orthonormal bases into each other are defined by~\cite{Schwinger.60}
\begin{equation}
\begin{aligned}
B_1 \leftrightarrow B_2:& \quad  {\bf f}_j =  \textsf{\textbf{U}}^{12}_{ji} {\bf e}_i,\textrm{ and }    {\bf e}_j =  {\left(\textsf{\textbf{U}}^{12}\right)^\dagger}_{ji} {\bf f}_i             \\
B_2 \leftrightarrow  B_3:& \quad  {\bf g}_k = \sum_{i=1}^2 \textsf{\textbf{U}}^{23}_{kj} {\bf f}_j, \textrm{ and }  {\bf f}_k = \sum_{i=1}^2 {\left(\textsf{\textbf{U}}^{23}\right)^\dagger}_{kj} {\bf g}_j     \\
B_1 \leftrightarrow  B_3:& \quad  {\bf g}_k = \sum_{i=1}^2 \textsf{\textbf{U}}^{13}_{ki} {\bf e}_i= \sum_{i,j=1}^2 \textsf{\textbf{U}}^{23}_{kj} \textsf{\textbf{U}}^{12}_{ji} {\bf e}_i, \\
& \textrm{ and }
{\bf e}_k= \sum_{i,j=1}^2 {\left(\textsf{\textbf{U}}^{12}\right)^\dagger}_{kj} {\left(\textsf{\textbf{U}}^{23}\right)^\dagger}_{ji} {\bf g}_i
.
\end{aligned}
\label{2020-hardy-eq-bt}
\end{equation}

One further ingredient of Hardy's configuration is a pure entangled state of two two-state particles
which can be parameterized by~\cite{Acin-2000}
(the relative order of states matter, therefore, as pointed out earlier, we shall omit a subscript referring to the first and second particle, respectively)
\begin{equation}
\begin{split}
\vert \Psi \rangle = \alpha \vert + + \rangle - \beta \vert - - \rangle,\textrm{ with }  \alpha, \beta \in \mathbb{R},\textrm{ and }           \\
\alpha^2 + \beta^2 = \cos^2 \phi + \sin^2 \phi = 1\textrm{ with } 0\le \phi \le \pi/4.
\end{split}
\label{2020-hardy-eq-es}
\end{equation}
The minus sign (indicating a phase $\varphi = \pi $ for which $e^{i \varphi}=-1$) has been chosen for the sake of conforming to Hardy's conventions.
Note that $0\le \alpha,\beta \le 1$.

So far, the transformation matrices in ~(\ref{2020-hardy-eq-bt}) have not yet been specified, but in order for
the argument to work, they should yield a faithful orthogonal representation of the orthogonality hypergraph
depicted in Fig.~\ref{2020-hardy-fig1}(a): in particular, one needs to assure that
\begin{equation}
\langle \Psi \vert uu \rangle =
\langle \Psi \vert vd \rangle =
\langle \Psi \vert dv \rangle  = 0.
\label{2020-hardy-eq-ortho}
\end{equation}
At the same time, and in order to obtain a contradiction with the classical prediction
``if $\Psi$ is true then $dd$ must be false''
or, in physical terms,
``if a system is prepared or (pre)selected in state $\Psi$ then an event or outcome associated with $dd$ cannot occur''
 (and vice versa), one needs to define those transformations such that,
in addition to~(\ref{2020-hardy-eq-ortho}),
\begin{equation}
\langle \Psi \vert dd \rangle \neq 0\textrm{ and ``as great as possible''.}
\label{2020-hardy-eq-contrad}
\end{equation}
In order to facilitate these desiderata~(\ref{2020-hardy-eq-ortho}) and (\ref{2020-hardy-eq-contrad}),
suppose {\it ad hoc} that
\begin{equation}
\begin{aligned}
\left(\textsf{\textbf{U}}^{12}\right)^\dagger& =  -\frac{i}{\sqrt{\alpha +\beta }}
\begin{pmatrix}
\sqrt{\beta} &  \sqrt{\alpha}\\
\sqrt{\alpha} & -\sqrt{\beta}
\end{pmatrix},
\\
\left(\textsf{\textbf{U}}^{23}\right)^\dagger &= \frac{1}{\sqrt{1 - \alpha \beta}}
\begin{pmatrix}
 \sqrt{\alpha \beta} &   -  \alpha  +\beta   \\
 \alpha   -  \beta  &  \sqrt{\alpha \beta}
\end{pmatrix},
\\
\left(\textsf{\textbf{U}}^{13}\right)^\dagger &=   \left(\textsf{\textbf{U}}^{12}\right)^\dagger \cdot \left(\textsf{\textbf{U}}^{23}\right)^\dagger
.
\end{aligned}
\label{2020-hardy-eq-uhardy}
\end{equation}

Assume further, without loss of generality, that  the first basis $B_1$ in~(\ref{2020-hardy-eq-b}) is identified with the Cartesian basis; that is,
$\vert + \rangle = \begin{pmatrix} 1,0 \end{pmatrix}$ and
$\vert - \rangle = \begin{pmatrix} 0,1 \end{pmatrix}$.
Consequently, the vectors of the other bases $B_2$ and $B_3$ are obtained by applying the respective transformations~(\ref{2020-hardy-eq-bt})
and~(\ref{2020-hardy-eq-uhardy}):
\begin{equation}
\begin{aligned}
\vert u \rangle  &=
\frac{i\begin{pmatrix}
\sqrt[4]{1-\alpha ^2},\sqrt{\alpha }
\end{pmatrix}}{\sqrt{\sqrt{1-\alpha ^2}+\alpha }}
,\\
\vert v \rangle  &=
\frac{i\begin{pmatrix}
\sqrt{\alpha } ,- \sqrt[4]{1-\alpha ^2}
\end{pmatrix}}{\sqrt{\sqrt{1-\alpha ^2}+\alpha }}
,\\
\vert c \rangle  &=
\frac{i\begin{pmatrix}
 \alpha ^{3/2}, \left(1-\alpha ^2\right)^{3/4}
\end{pmatrix}}{\sqrt{\alpha ^3-\sqrt{1-\alpha ^2} \alpha ^2+\sqrt{1-\alpha ^2}}}
,\\
\vert d \rangle  &=
\frac{i\begin{pmatrix}
 \left(1-\alpha ^2\right)^{3/4} ,  -\alpha ^{3/2}
\end{pmatrix}    }{\sqrt{\alpha ^3-\sqrt{1-\alpha ^2} \alpha ^2+\sqrt{1-\alpha ^2}}}
.
\end{aligned}
\label{2020-hardy-eq-bvrep}
\end{equation}
We are only dealing with pure states represented as normalized
vectors which are (the sum of) the Kronecker (that is, ``delineated'' outer or tensor) products~\cite[Chapter~1]{mermin-07}, e.g.,
$\vert \Psi \rangle =
\alpha \begin{pmatrix}1,0\end{pmatrix}\otimes \begin{pmatrix}1,0\end{pmatrix}    -\sqrt{1-\alpha ^2}\begin{pmatrix}0,1\end{pmatrix} \otimes \begin{pmatrix}0,1\end{pmatrix} =
\begin{pmatrix}\alpha ,0,0,-\sqrt{1-\alpha ^2}\end{pmatrix}$.
The associated propositional observables can then be written in terms of the orthogonal projections formed
as dyadic products $\vert x \rangle \langle x \vert$ of the unit (state) vectors
$\vert x \rangle$.

\begin{widetext}
By applying the
transformations~(\ref{2020-hardy-eq-bt}),
$\vert \Psi \rangle $ can be rewritten in terms of either
(i) the second basis $B_2$ for the first particle, and the second basis $B_2$ for the second particle, or
(ii) the second basis $B_2$ for the first particle, and the third basis $B_3$ for the second particle, or
(iii) the third basis $B_3$ for the first particle, and the second basis $B_2$ for the second particle, or
(iv) the third basis $B_3$ for the first particle, and the third basis $B_3$ for the second particle, that is,
\begin{eqnarray}
\vert \Psi \rangle
&=&
- \vert u  v \rangle  \sqrt{\alpha  \beta }- \vert v  \rangle  \left( \vert u  \rangle  \sqrt{\alpha  \beta
   }+ \vert v  \rangle  (\alpha -\beta )\right),
\label{2020-hardy-eq-uhardy1}  \\
&=&
\frac{1}{\sqrt{1-\alpha  \beta }}\left[    \vert u  c  \rangle  \left(\sqrt{\alpha  \beta ^3}-\sqrt{\alpha ^3 \beta
   }\right)- \vert v  c \rangle    \left(\alpha ^2-\alpha  \beta +\beta ^2\right)-
     \vert u  d \rangle \alpha  \beta    \right], \label{2020-hardy-eq-uhardy2}   \\
&=&
\frac{1 }{\sqrt{1-\alpha  \beta }} \left[\vert c  u  \rangle  \left(\sqrt{\alpha  \beta ^3}-\sqrt{\alpha ^3 \beta
   }\right)- \vert c  v  \rangle  \left(\alpha ^2-\alpha  \beta +\beta ^2\right)-
      \vert d  u \rangle  \alpha  \beta  \right],   \label{2020-hardy-eq-uhardy3}   \\
&=&  -
\frac{1}{1 - \alpha  \beta}\left[ \vert c c  \rangle  (\alpha -\beta ) \left(\alpha ^2+\beta ^2\right)+ \left(\vert c d \rangle
+ \vert d  c  \rangle  \right)  (\alpha  \beta )^{3/2}+
     \vert d  d  \rangle \alpha  \beta    (\beta -\alpha )\right]. \label{2020-hardy-eq-uhardy4}
\end{eqnarray}
\end{widetext}
As can be readily read off from these representations of $\vert \Psi \rangle$ the conditions~(\ref{2020-hardy-eq-ortho})
and desideratum~(\ref{2020-hardy-eq-contrad}) are satisfied:
(\ref{2020-hardy-eq-uhardy1}) has no term proportional to $\vert uu \rangle$,
(\ref{2020-hardy-eq-uhardy2}) has no term proportional to $\vert vd \rangle$,
(\ref{2020-hardy-eq-uhardy3}) has no term proportional to $\vert dv \rangle$, and
(\ref{2020-hardy-eq-uhardy4}) has a term proportional to $\vert dd \rangle$.

To complete Hardy's original argument  we compare the classical prediction of ``zero outcome'' (nonoccurrence)
for observable $dd$
to the quantum prediction probability
\begin{equation}
\vert \langle  dd \vert \Psi \rangle \vert^2 =
\left\{
\frac{\alpha  \left[\alpha  \left(\sqrt{1-\alpha ^2}+\alpha \right)-1\right] }{ \alpha  \sqrt{1-\alpha ^2}-1  }
\right\}^2
\label{2020-hardy-eq-qprobPsidd}
\end{equation}
obtained from preparing (also known as preselecting) two entangled particles in state $\vert \Psi \rangle$ and measuring (e.g. by postselection) the
nonvanishing probability to find them in state $\vert dd \rangle$ (thus contradicting aforementioned classical predictions).
$\vert \langle  dd \vert \Psi \rangle \vert^2 $ acquires its maximal value $\frac{1}{2} \left(5 \sqrt{5}-11\right)\approx 0.09$ at
$\alpha_\pm = \sqrt{1 \pm \sqrt{6 \sqrt{5}-13}}/\sqrt{2} $.
This is
slightly below the maximal violation of the three-dimensional ``minimal'' true-implies-false case
[the Specker bug~\cite{kochen2,2018-minimalYIYS,svozil-2017-b} depicted in Fig.~\ref{2020-hardy-fig1}(b)]
with probability  $1/9\approx 0.1$~\cite{Belinfante-73,redhead,cabello-1994,Cabello-1996-diss,svozil-tkadlec}.

\section{Varieties of coordinatization}

In what follows, we shall enumerate a few faithful orthogonal representations of the Hardy gadget.
Presently, no general analytic construction for finding even a single faithful orthogonal representation of a (hyper)graph (if any) exists,
let alone a method for finding all such coordinatizations.
Nevertheless, {\it ad hoc} faithful orthogonal representations can be generated in extenso by heuristic algorithms.
With regards to (in)decomposability, the Hardy gadget allows almost all types of faithful orthogonal representations:
``mixed'' ones which have entangled as well as factorizable states,
and ones which use entangled states.
Entirely decomposable configurations are prohibited for geometric reasons.

From now on the observables need not be formed by some sort of composition,
and therefore two symbols such as ``$uv$'' should only be understood as a label.
Note that indecomposable vectors can be interpreted as pure entangled states.
Likewise decomposable vectors represent pure factorizable states.
The coordinatizations will not be enumerated completely, as only the intertwining vertices will be explicitly mentioned.
Nevertheless, completions are straightforward and have been discussed earlier.
Reference~\cite{havlicek-svozil-2020-dec} contains a careful categorization with respect to (in)decomposability.

\subsection{Mixed (in)decomposabability}

Previous parametrizations~\cite{Hardy-92,Hardy-93,cabello-96,cabello-97-nhvp,Badziag-2011} of Hardy's  (minimal with respect to the number of vertices in four dimensions~\cite{2018-minimalYIYS})
true-implies-false gadget depicted in Fig.~\ref{2020-hardy-fig1}(a) in terms of four-dimensional  vectors
appear to be motivated by high yield -- that is, by maximizing the quantum predictions of occurrence of the output (postselection) port $dd$,
as well as by (in)decomposability of the associated vectors.
This is motivated by what is sometimes referred to as ``demonstrations of nonlocal contextuality''; that is, the ``spread'' of the relational information~\cite{zeil-99} among
pairs of (spacelike) separated particles.

The first, mixed with respect to (in)decomposability of the vectors,
type of coordinatization can almost directly be read off from the orthogonality hypergraph of the Hardy gadget depicted in Fig.~\ref{2020-hardy-fig1}(a).
Note that the two ``central full contexts''
$\{\vert cv \rangle ,\vert vu \rangle ,\vert uu \rangle ,\vert dv \rangle \}$
and
$\{\vert vc \rangle ,\vert uv \rangle ,\vert uu \rangle ,\vert vd \rangle \}$
intertwine at one common element $\vert uu \rangle$ and
are actually ``generated'' by the flattened tensor products of two nonidentical two-dimensional contexts
representable by the two orthonormal bases
$\{\vert c \rangle ,\vert d \rangle\}$ and
$\{\vert u \rangle ,\vert v \rangle\}$, respectively.
Hence all that is necessary is to make sure that $\vert \Psi \rangle$ is orthogonal to three vectors
$\vert vd \rangle$,
$\vert uu \rangle$, and
$\vert dv \rangle$
of four-dimensional space (and no multiplicities occur), as already encoded in Eqs.~(\ref{2020-hardy-eq-ortho}):
\begin{equation}
\begin{aligned}
\vert \Psi \rangle
\propto
\big(
&d_2 u_2^2 v_1 - 2 d_2 u_1 u_2 v_2 + d_1 u_2^2 v_2,\\
&d_2 u_1^2 v_2 - d_1 u_2^2 v_1 , \\
& d_2 u_1^2 v_2 - d_1 u_2^2 v_1   ,\\
&2 d_1 u_1 u_2 v_1 - d_2 u_1^2 v_1  - d_1 u_1^2 v_2
\big)
,
\end{aligned}
\label{2020-hardy-eq-psigeneral}
\end{equation}
where $x_i$ stands for the $i$th component of the vector $x$ with respect to some basis common to all vectors.

In order to be able to claim nonlocality,
additional constraints can be required from the components of $\vert \Psi \rangle  $.
Suppose one desires $\vert \Psi \rangle$ to be entangled.
Then the product of its outer components should not be equal to the product
of its inner components;
that is, $\Psi_1 \Psi_4 \neq \Psi_2 \Psi_3$
because every decomposable product state of two vectors with components
$\begin{pmatrix}a,b\end{pmatrix}$ and
$\begin{pmatrix}c,d\end{pmatrix}$
is of the (delineated) form
$\begin{pmatrix}x_1= ac,x_2=ad,x_3=bc,x_4=bd\end{pmatrix}$, so that, because of commutativity of scalars,
$x_1 x_4 = (ac)(bd) = abcd = (ad)(bc) =x_2 x_3$.
If one prefers the tensor product in matrix notation then
$\begin{pmatrix}x_1= ac&x_2=ad\\x_3=bc&x_4=bd\end{pmatrix}$
and the criterion for factorizability and decomposability is a vanishing determinant
$x_1 x_4 - x_2 x_3 = 0$.
Applying this constraint to Eq.~(\ref{2020-hardy-eq-psigeneral}) results in
\begin{equation}
(d_2 u_1 - d_1 u_2) (u_1 v_2 - u_2 v_1) \neq 0
.
\end{equation}
The third and the fourth rows of Table~\ref{2020-Hardy-FOR} containing vectors present two {\it ad hoc}
configurations satisfying this ``indecomposability'' constraint.

\subsection{Indecomposable configurations}

The last two rows of Table~\ref{2020-Hardy-FOR} contain faithful orthogonal
representations of the Hardy gadget in which all intertwining vectors are entangled
(e.g., in the complex realization because the number of components of any vector with imaginary units $i$ and $-i$ is odd;
that is, either one or three).
These coordinatizations have been obtained with a heuristic algorithm developed by Mc{K}ay, Megill and Pavi{\v{c}}i{\'{c}}~\cite{Pavii2018}.


 \begin{table*}
\centering
 \caption{\label{2020-Hardy-FOR}  Tabulation of some faithful orthogonal representations of the Hardy gadget.
Two missing vectors per context as well as normalizations can be completed with a little effort.
Labels of the form ``$ab$'' should not be understood as product states but have been used merely to conform to Hardy's original nomenclature.}
\begin{ruledtabular}
 \begin{tabular}{ccccccccccccccccccccccc}
    &{$\psi$}&{$dv$}&{$vd$}&{$uu$}&{$uv$}&{$vu$}&{$cv$}&{$vc$}&{$dd$}\\
\hline
{\scriptsize CEG-A 1996~\cite{cabello-96}}
&
{\scriptsize $\begin{pmatrix}1,-1,-1,0\end{pmatrix}$}&
{\scriptsize $\begin{pmatrix}1,0,1,0\end{pmatrix}$}&
{\scriptsize $\begin{pmatrix}1,1,0,0\end{pmatrix}$}&
{\scriptsize $\begin{pmatrix}0,0,0,1\end{pmatrix}$}&
{\scriptsize $\begin{pmatrix}0,0,1,0\end{pmatrix}$}&
{\scriptsize $\begin{pmatrix}0,1,0,0\end{pmatrix}$}&
{\scriptsize $\begin{pmatrix}1,0,-1,0\end{pmatrix}$}&
{\scriptsize $\begin{pmatrix}1,-1,0,0\end{pmatrix}$}&
{\scriptsize $\begin{pmatrix}1,1,1,1\end{pmatrix}$}
\\
{\scriptsize Cabello 1997~\cite{cabello-97-nhvp}}&
{\scriptsize $AB$}&
{\scriptsize $\beta_+$}&
{\scriptsize $\beta-$}&
{\scriptsize $\alpha$}&
{\scriptsize $\delta_-$}&
{\scriptsize $\delta_+$}&
{\scriptsize $\gamma_+$}&
{\scriptsize $\gamma_-$}&
{\scriptsize $aB$}
\\
{\tiny $A=B=\begin{pmatrix} 1,0\end{pmatrix}$},
{\tiny $a=\begin{pmatrix}\frac{1}{\sqrt{3}},\frac{-2\sqrt{2}}{\sqrt{3}}\end{pmatrix}$}&
{\scriptsize $\begin{pmatrix}1,0,0,0\end{pmatrix}$}&
{\scriptsize $\begin{pmatrix}0,\frac{1}{2},\frac{-\sqrt{3}}{2},0\end{pmatrix}$}&
{\scriptsize $\begin{pmatrix}0,\frac{1}{2},\frac{\sqrt{3}}{2},0\end{pmatrix}$}&
{\scriptsize $\begin{pmatrix}0,0,0,1\end{pmatrix}$}&
{\tiny $\begin{pmatrix}\frac{1}{\sqrt{3}},\frac{1}{\sqrt{2}},\frac{-1}{\sqrt{6}},0\end{pmatrix}$}&
{\tiny $\begin{pmatrix}\frac{-1}{\sqrt{3}},\frac{1}{\sqrt{2}},\frac{1}{\sqrt{6}},0\end{pmatrix}$}&
{\tiny $\begin{pmatrix}\sqrt{\frac{2}{3}},\frac{1}{2},\frac{1}{2\sqrt{3}},0\end{pmatrix}$}&
{\tiny $\begin{pmatrix}\sqrt{\frac{2}{3}},\frac{-1}{2},\frac{1}{2\sqrt{3}},0\end{pmatrix}$}&
{\tiny $\begin{pmatrix}\frac{1}{3},0,\frac{-2\sqrt{2}}{3},0\end{pmatrix}$}
\\
{\scriptsize BBCGL 2011~\cite{Badziag-2011}}&
{\scriptsize $\Psi$}&
{\scriptsize $\bar{a}_2b_1$}&
{\scriptsize $a_1\bar{b}_2$}&
{\scriptsize $a_2b_2$}&
{\scriptsize $\bar{a}_2b_2$}&
{\scriptsize $a_2\bar{b}_2$}&
{\scriptsize $\bar{a}_2\bar{b}_1$}&
{\scriptsize $\bar{a}_1\bar{b}_2$}&
{\scriptsize $a_1b_1$}
\\
{\scriptsize $u=\begin{pmatrix}1,0\end{pmatrix}$, $c=\begin{pmatrix}1,1\end{pmatrix}$}&
{\scriptsize $\begin{pmatrix}0, -1, -1, -1\end{pmatrix}$}&
{\scriptsize $\begin{pmatrix}0, 1, 0, -1\end{pmatrix}$}&
{\scriptsize $\begin{pmatrix}0, 0, 1, -1\end{pmatrix}$}&
{\scriptsize $\begin{pmatrix}1, 0, 0, 0\end{pmatrix}$}&
{\scriptsize $\begin{pmatrix}0, 1, 0, 0\end{pmatrix}$}&
{\scriptsize $\begin{pmatrix}0, 0, 1, 0\end{pmatrix}$}&
{\scriptsize $\begin{pmatrix}0, 1, 0, 1\end{pmatrix}$}&
{\scriptsize $\begin{pmatrix}0, 0, 1, 1\end{pmatrix}$}&
{\scriptsize $\begin{pmatrix}1, -1, -1, 1\end{pmatrix}$}
\\
{\scriptsize $u=\begin{pmatrix}1,0\end{pmatrix}$, $c=\begin{pmatrix}2,3\end{pmatrix}$}&
{\scriptsize $\begin{pmatrix}0, -2, -2, -3\end{pmatrix}$}&
{\scriptsize $\begin{pmatrix}0, 3, 0, -2\end{pmatrix}$}&
{\scriptsize $\begin{pmatrix}0, 0, 3, -2\end{pmatrix}$}&
{\scriptsize $\begin{pmatrix}1, 0, 0, 0\end{pmatrix}$}&
{\scriptsize $\begin{pmatrix}0, 1, 0, 0\end{pmatrix}$}&
{\scriptsize $\begin{pmatrix}0, 0, 1, 0\end{pmatrix}$}&
{\scriptsize $\begin{pmatrix}0, 2, 0, 3\end{pmatrix}$}&
{\scriptsize $\begin{pmatrix}0, 0, 2, 3\end{pmatrix}$}&
{\scriptsize $\begin{pmatrix}9, -6, -6, 4\end{pmatrix}$}
\\
%
%
{\scriptsize {\tt VECFIND}~\cite{Pavii2018} $\{0,1,-2,\sqrt{2}\}$}
&
{\scriptsize $\begin{pmatrix}1,-2,\sqrt{2},0\end{pmatrix}$}&
{\scriptsize $\begin{pmatrix}-2,0,\sqrt{2},0\end{pmatrix}$}&
{\scriptsize $\begin{pmatrix}0,1,\sqrt{2},0\end{pmatrix}$}&
{\scriptsize $\begin{pmatrix}0,0,0,1\end{pmatrix}$}&
{\scriptsize $\begin{pmatrix}1,0,0,0\end{pmatrix}$}&
{\scriptsize $\begin{pmatrix}0,1,0,0\end{pmatrix}$}&
{\scriptsize $\begin{pmatrix}1,0,\sqrt{2},0\end{pmatrix}$}&
{\scriptsize $\begin{pmatrix}0,-2,\sqrt{2},0\end{pmatrix}$}&
{\scriptsize $\begin{pmatrix}-2,1,\sqrt{2},0\end{pmatrix}$}
\\
{\scriptsize {\tt VECFIND} $\{1,2,\frac{-1}{2},3,5,\pm i\}$}
&
{\scriptsize $\begin{pmatrix}i,3,3,5\end{pmatrix}$}&
{\scriptsize $\begin{pmatrix}i,\frac{-1}{2},1,\frac{-1}{2}\end{pmatrix}$}&
{\scriptsize $\begin{pmatrix}i,1,\frac{-1}{2},\frac{-1}{2}\end{pmatrix}$}&
{\scriptsize $\begin{pmatrix}1,i,i,-i\end{pmatrix}$}&
{\scriptsize $\begin{pmatrix}1,i,-i,i\end{pmatrix}$}&
{\scriptsize $\begin{pmatrix}1,-i,i,i\end{pmatrix}$}&
{\scriptsize $\begin{pmatrix}i,2,1,2\end{pmatrix}$}&
{\scriptsize $\begin{pmatrix}i,1,2,2\end{pmatrix}$}&
{\scriptsize $\begin{pmatrix}5,i,i,i\end{pmatrix}$}
\\
{\scriptsize {\tt VECFIND} $\{1,-1,2,\frac{1}{2},3,5\}$}
&
{\scriptsize $\begin{pmatrix}-1,3,3,5                     \end{pmatrix}$}&
{\scriptsize $\begin{pmatrix}1,-1,\frac{1}{2},\frac{1}{2} \end{pmatrix}$}&
{\scriptsize $\begin{pmatrix}1,1,1,-1                     \end{pmatrix}$}&
{\scriptsize $\begin{pmatrix}1,-1,1,1                     \end{pmatrix}$}&
{\scriptsize $\begin{pmatrix}-1,2,1,2                     \end{pmatrix}$}&
{\scriptsize $\begin{pmatrix}1,-1,-1,2                    \end{pmatrix}$}&
{\scriptsize $\begin{pmatrix}-1,1,2,2                     \end{pmatrix}$}&
{\scriptsize $\begin{pmatrix}1,1,-1,1                     \end{pmatrix}$}&
{\scriptsize $\begin{pmatrix}1,\frac{1}{2},-1,\frac{1}{2} \end{pmatrix}$}
\end{tabular}

\end{ruledtabular}
\end{table*}

\subsection{Impossibility of complete decomposability}

Conversely, it might be desirable to keep $\vert \Psi \rangle$ decomposable and factorizable; that is, all entities should be in a product state.
In this case, the product of the outer components of $\vert \Psi \rangle$ should  be equal to the product  of its inner components;
that is, $\Psi_1 \Psi_4 = \Psi_2 \Psi_3$.
This results in the constraint $d_1 = d_2 u_1/u_2$,  with  $u_2 \neq 0$ from Eq.~(\ref{2020-hardy-eq-psigeneral}),
and consequently in a multiplicity of vectors; more explicitly, $d \propto u$.
The resulting unattainability of a coordinatization with purely decomposable vectors should come as no surprise
as  the hypergraph of the Hardy gadget depicted in Fig.~\ref{2020-hardy-fig1}(a) contains three triangular subgraphs,
namely the cyclically intertwining sets of contexts
$\{ \{ \Psi,2,3,vd \} , \{ vd,uv,vc,uu \} , \{ uu,20,21,\Psi \} \}$,
$\{ \{ \Psi,16,17,dv \} , \{ dv,vu,cv,uu \} , \{ uu,20,21,\Psi \} \}$, as well as
$\{  \{ vu,dv,cv,uu \} ,\{ uu,vd,vc,uv \}  ,  \{ vu,18,19,uv \} \}$.
Triangular hypergraphs
have no faithful orthogonal representation by purely decomposable vectors~\cite{havlicek-svozil-2020-dec}.

\section{Extensions to true-implies-true gadgets}

We now turn to important extensions of the Hardy gadget which have a classical true-implies-true structure, as already
employed in Kochen and Specker's $\Gamma_1$~\cite{kochen1}
and discussed in Ref.~\cite{2018-minimalYIYS}.
A further escalation is a combination of these true-implies-true gadgets,
similar to Kochen and Specker's $\Gamma_3$, which delivers a truly nonclassical performance on the algebraic level
of the propositional observables (and not just probabilistic predictions based upon classical probabilities):
Unlike the Hardy and its extended true-implies-true gadgets,
those observables can no longer be faithfully embedded into any Boolean algebra~\cite[Theorem~0]{kochen1}.

\begin{figure}
\begin{center}
\begin{tabular}{ c c }
\resizebox{.25\textwidth}{!}{%
\begin{tikzpicture}  [scale=0.5]

\tikzstyle{every path}=[line width=1pt]

\newdimen\ms
\ms=0.1cm
\tikzstyle{s1}=[color=red,rectangle,inner sep=3.5]
\tikzstyle{c4}=[circle,inner sep={\ms/8},minimum size=5*\ms]
\tikzstyle{c3}=[circle,inner sep={\ms/8},minimum size=4*\ms]
\tikzstyle{c2}=[circle,inner sep={\ms/8},minimum size=3*\ms]
\tikzstyle{c1}=[circle,inner sep={\ms/8},minimum size=2*\ms]
\tikzstyle{cs1}=[circle,inner sep={\ms/8},minimum size=1*\ms]


\coordinate (psi) at (1,0);
\coordinate (vd) at (0,2);
\coordinate (dv) at (4,2);
\coordinate (uu) at (2,4);
\coordinate (vu) at (1,5);
\coordinate (uv) at (3,5);
\coordinate (cv) at (0,6);
\coordinate (vc) at (4,6);
\coordinate (dd) at (2.5,8);

\coordinate (O) at (4.5,1);
\coordinate (P) at (8,8);

\coordinate (M) at (6.5,6);
\coordinate (N) at (8,0);


\draw [color=orange] (psi) -- (vd)  coordinate[cs1,fill=white,draw=gray,pos=0.33,label=below left:{\scriptsize \color{gray}2}] (2)  coordinate[cs1,fill=white,draw=gray,pos=0.66,label=below left:{\scriptsize \color{gray}3}] (3);
\draw [color=blue] (psi) -- (uu)  coordinate[cs1,fill=white,draw=gray,pos=0.33,label={[left,xshift=0.2mm]:{\scriptsize \color{gray}20}}] (20)  coordinate[cs1,fill=white,draw=gray,pos=0.66,label={[left,xshift=0.2mm]:{\scriptsize \color{gray}21}}] (21);
\draw [color=red] (psi) -- (dv) coordinate[cs1,fill=white,draw=gray,pos=0.33,label=above:{\scriptsize \color{gray}17}]  (17) coordinate[cs1,fill=white,draw=gray,pos=0.66,label=above:{\scriptsize \color{gray}16}] (16);
\draw [color=green] (vd) -- (vc);
\draw [color=gray] (dv) -- (cv);
\draw [color=magenta] (vu) -- (uv) coordinate[cs1,fill=white,draw=gray,pos=0.33,label=above:{\scriptsize \color{gray}18}] (18)  coordinate[cs1,fill=white,draw=gray,pos=0.66,label=above:{\scriptsize \color{gray}19}] (19);
\draw [color=cyan] (cv) -- (dd) coordinate[cs1,fill=white,draw=gray,pos=0.33,label=above left:{\scriptsize \color{gray}8}]  (8) coordinate[cs1,fill=white,draw=gray,pos=0.66,label=above left:{\scriptsize \color{gray}9}] (9);
\draw [color=olive] (vc) -- (dd) coordinate[cs1,fill=white,draw=gray,pos=0.33,label=right:{\scriptsize \color{gray}12}] (12)  coordinate[cs1,fill=white,draw=gray,pos=0.66,label=below:{\scriptsize \color{gray}11}] (11);

\draw [ForestGreen] plot [smooth] coordinates {(psi)  (O) (M) (P)};
\draw [RubineRed] plot [smooth] coordinates {(uu) (dd) (M) (N)};


\draw (psi) coordinate[c4,fill=orange,label=below:$\Psi$];
\draw (psi) coordinate[c3,fill=blue];
\draw (psi) coordinate[c2,fill=red];
\draw (psi) coordinate[c1,fill=ForestGreen];

\draw (vd) coordinate[c2,fill=orange,label=left:$vd$];
\draw (vd) coordinate[c1,fill=green];

\draw (dv) coordinate[c2,fill=red,label=above:$dv$];
\draw (dv) coordinate[c1,fill=gray];

\draw (uu) coordinate[c4,fill=gray,label=right:$uu$];
\draw (uu) coordinate[c3,fill=green];
\draw (uu) coordinate[c2,fill=RubineRed];
\draw (uu) coordinate[c1,fill=blue];

\draw (vu) coordinate[c2,fill=gray,label=left:$vu$];
\draw (vu) coordinate[c1,fill=magenta];

\draw (uv) coordinate[c2,fill=green,label=right:$uv$];
\draw (uv) coordinate[c1,fill=magenta];

\draw (cv) coordinate[c2,fill=gray,label=left:$cv$];
\draw (cv) coordinate[c1,fill=cyan];

\draw (vc) coordinate[c2,fill=green,label=right:$vc$];
\draw (vc) coordinate[c1,fill=olive];

\draw (dd) coordinate[c3,fill=olive,label=above:$dd$];
\draw (dd) coordinate[c2,fill=cyan];
\draw (dd) coordinate[c1,fill=RubineRed];

\draw (N) coordinate[c2,fill=RubineRed,label=right:$N$];

\draw (M) coordinate[c2,fill=RubineRed,label=right:$M$];
\draw (M) coordinate[c1,fill=ForestGreen];

\draw (O) coordinate[c2,fill=white,draw=gray,label=right:{\scriptsize \color{gray}$O$}];

\draw (P) coordinate[c2,fill=white,draw=gray,label=right:{\scriptsize \color{gray}$P$}];
\end{tikzpicture}
}
&
\resizebox{.22\textwidth}{!}{%
\begin{tikzpicture}  [scale=0.5]

\tikzstyle{every path}=[line width=1pt]

\newdimen\ms
\ms=0.1cm
\tikzstyle{s1}=[color=red,rectangle,inner sep=3.5]
\tikzstyle{c3}=[circle,inner sep={\ms/8},minimum size=4*\ms]
\tikzstyle{c2}=[circle,inner sep={\ms/8},minimum size=3*\ms]
\tikzstyle{c1}=[circle,inner sep={\ms/8},minimum size=2*\ms]
\tikzstyle{cs1}=[circle,inner sep={\ms/8},minimum size=1*\ms]


\coordinate (a1) at (2,0);
\coordinate (a2) at (0,2);
\coordinate (a3) at (4,2);
\coordinate (uu) at (2,4);
\coordinate (a4) at (1,5);
\coordinate (a5) at (3,5);
\coordinate (a6) at (0,6);
\coordinate (a7) at (4,6);
\coordinate (a8) at (2,8);

\coordinate (O) at (7,8);
\coordinate (M) at (6,4);
\coordinate (N) at (7,0);

\coordinate (aux1) at (4.5,0.5);
\coordinate (aux2) at (4.5,7.5);


\draw [color=orange] (a1) -- (a2)  coordinate[cs1,fill=white,draw=gray,pos=0.5] (2) ;
\draw [color=red] (a1) -- (a3) coordinate[cs1,fill=white,draw=gray,pos=0.5]  (17);
\draw [color=green] (a2) -- (a7);
\draw [color=gray] (a3) -- (a6);
\draw [color=magenta] (a4) -- (a5) coordinate[cs1,fill=white,draw=gray,pos=0.5] (18);
\draw [color=cyan] (a6) -- (a8) coordinate[cs1,fill=white,draw=gray,pos=0.5]  (8);
\draw [color=olive] (a7) -- (a8) coordinate[cs1,fill=white,draw=gray,pos=0.5] (12);

\draw [ForestGreen] plot [smooth] coordinates {(a1) (aux1) (M) (O)};
\draw [RubineRed] plot [smooth] coordinates {(a8) (aux2) (M) (N)};


\draw (a1) coordinate[c3,fill=orange,label=below:$a_1$];
\draw (a1) coordinate[c2,fill=ForestGreen];
\draw (a1) coordinate[c1,fill=red];

\draw (a2) coordinate[c2,fill=orange,label=left:$a_2$];
\draw (a2) coordinate[c1,fill=green];

\draw (a3) coordinate[c2,fill=red,label=above:$a_3$];
\draw (a3) coordinate[c1,fill=gray];

\draw (a4) coordinate[c2,fill=gray,label=left:$a_4$];
\draw (a4) coordinate[c1,fill=magenta];

\draw (a5) coordinate[c2,fill=green,label=right:$a_5$];
\draw (a5) coordinate[c1,fill=magenta];

\draw (a6) coordinate[c2,fill=gray,label=left:$a_6$];
\draw (a6) coordinate[c1,fill=cyan];

\draw (a7) coordinate[c2,fill=green,label=above:$a_7$];
\draw (a7) coordinate[c1,fill=olive];

\draw (a8) coordinate[c3,fill=olive,label=above:$a_8$];
\draw (a8) coordinate[c2,fill=cyan];
\draw (a8) coordinate[c1,fill=RubineRed];

\draw (N) coordinate[c2,fill=RubineRed,label=right:$N$];

\draw (M) coordinate[c2,fill=RubineRed,label=right:$M$];
\draw (M) coordinate[c1,fill=ForestGreen];

\draw (O) coordinate[c2,fill=white,draw=gray,label=right:{\scriptsize \color{gray}$O$}];

\end{tikzpicture}
}
\\
(a) & (b)
\\

\resizebox{.25\textwidth}{!}{%
\begin{tikzpicture}  [scale=0.5]

\tikzstyle{every path}=[line width=1pt]

\newdimen\ms
\ms=0.1cm
\tikzstyle{s1}=[color=red,rectangle,inner sep=3.5]
\tikzstyle{c3}=[circle,inner sep={\ms/8},minimum size=4*\ms]
\tikzstyle{c2}=[circle,inner sep={\ms/8},minimum size=3*\ms]
\tikzstyle{c1}=[circle,inner sep={\ms/8},minimum size=2*\ms]
\tikzstyle{cs1}=[circle,inner sep={\ms/8},minimum size=1*\ms]


\coordinate (psi) at (1,0);
\coordinate (vd) at (0,2);
\coordinate (dv) at (4,2);
\coordinate (uu) at (2,4);
\coordinate (vu) at (1,5);
\coordinate (uv) at (3,5);
\coordinate (cv) at (0,6);
\coordinate (vc) at (4,6);
\coordinate (dd) at (2.5,8);

\coordinate (O) at (4.5,1);
\coordinate (P) at (8,8);

\coordinate (M) at (6.5,6);
\coordinate (N) at (8,0);


\draw [color=lightgray] (psi) -- (vd)  coordinate[cs1,fill=white,draw=lightgray,pos=0.33,label=below left:{\scriptsize \color{lightgray}2}] (2)  coordinate[cs1,fill=white,draw=lightgray,pos=0.66,label=below left:{\scriptsize \color{lightgray}3}] (3);
\draw [color=lightgray] (psi) -- (uu)  coordinate[cs1,fill=white,draw=lightgray,pos=0.33,label={[left,xshift=0.2mm]:{\scriptsize \color{lightgray}20}}] (20)  coordinate[cs1,fill=white,draw=lightgray,pos=0.66,label={[left,xshift=0.2mm]:{\scriptsize \color{lightgray}21}}] (21);
\draw [color=lightgray] (psi) -- (dv) coordinate[cs1,fill=white,draw=lightgray,pos=0.33,label=above:{\scriptsize \color{lightgray}17}]  (17) coordinate[cs1,fill=white,draw=lightgray,pos=0.66,label=above:{\scriptsize \color{lightgray}16}] (16);
\draw [color=lightgray] (vd) -- (vc);
\draw [color=lightgray] (dv) -- (cv);
\draw [color=lightgray] (vu) -- (uv) coordinate[cs1,fill=white,draw=lightgray,pos=0.33,label=above:{\scriptsize \color{lightgray}18}] (18)  coordinate[cs1,fill=white,draw=lightgray,pos=0.66,label=above:{\scriptsize \color{lightgray}19}] (19);
\draw [color=lightgray] (cv) -- (dd) coordinate[cs1,fill=white,draw=lightgray,pos=0.33,label=above left:{\scriptsize \color{lightgray}8}]  (8) coordinate[cs1,fill=white,draw=lightgray,pos=0.66,label=above left:{\scriptsize \color{lightgray}9}] (9);
\draw [color=lightgray] (vc) -- (dd) coordinate[cs1,fill=white,draw=lightgray,pos=0.33,label=right:{\scriptsize \color{lightgray}12}] (12)  coordinate[cs1,fill=white,draw=lightgray,pos=0.66,label=below:{\scriptsize \color{lightgray}11}] (11);

\draw [lightgray] plot [smooth] coordinates {(psi)  (O) (M) (P)};
\draw [lightgray] plot [smooth] coordinates {(uu) (dd) (M) (N)};


\draw (psi) coordinate[c2,fill=red,label=below:$\Psi$];

\draw (vd) coordinate[c1,fill=lightgray,label=left:$vd$];

\draw (dv) coordinate[c1,fill=lightgray,label=above:$dv$];

\draw (uu) coordinate[c1,fill=lightgray,label=right:$uu$];

\draw (vu) coordinate[c1,fill=lightgray,label=left:$vu$];

\draw (uv) coordinate[c1,fill=lightgray,label=right:$uv$];

\draw (cv) coordinate[c1,fill=lightgray,label=left:$cv$];

\draw (vc) coordinate[c1,fill=lightgray,label=right:$vc$];

\draw (dd) coordinate[c2,draw=green,fill=white,label=above:$dd$];

\draw (N) coordinate[c2,fill=red,label=right:$N$];

\draw (M) coordinate[c2,draw=green,fill=white,label=right:$M$];

\draw (O) coordinate[c2,draw=green,fill=white,label=right:{\scriptsize \color{lightgray}$O$}];

\draw (P) coordinate[c2,draw=green,fill=white,label=right:{\scriptsize \color{lightgray}$P$}];
\end{tikzpicture}
}
&
\resizebox{.22\textwidth}{!}{%
\begin{tikzpicture}  [scale=0.5]

\tikzstyle{every path}=[line width=1pt]

\newdimen\ms
\ms=0.1cm
\tikzstyle{s1}=[color=lightgray,rectangle,inner sep=3.5]
\tikzstyle{c3}=[circle,inner sep={\ms/8},minimum size=4*\ms]
\tikzstyle{c2}=[circle,inner sep={\ms/8},minimum size=3*\ms]
\tikzstyle{c1}=[circle,inner sep={\ms/8},minimum size=2*\ms]
\tikzstyle{cs1}=[circle,inner sep={\ms/8},minimum size=1*\ms]


\coordinate (a1) at (2,0);
\coordinate (a2) at (0,2);
\coordinate (a3) at (4,2);
\coordinate (uu) at (2,4);
\coordinate (a4) at (1,5);
\coordinate (a5) at (3,5);
\coordinate (a6) at (0,6);
\coordinate (a7) at (4,6);
\coordinate (a8) at (2,8);

\coordinate (O) at (7,8);
\coordinate (M) at (6,4);
\coordinate (N) at (7,0);

\coordinate (aux1) at (4.5,0.5);
\coordinate (aux2) at (4.5,7.5);


\draw [color=lightgray] (a1) -- (a2)  coordinate[cs1,fill=white,draw=lightgray,pos=0.5] (2) ;
\draw [color=lightgray] (a1) -- (a3) coordinate[cs1,fill=white,draw=lightgray,pos=0.5]  (17);
\draw [color=lightgray] (a2) -- (a7);
\draw [color=lightgray] (a3) -- (a6);
\draw [color=lightgray] (a4) -- (a5) coordinate[cs1,fill=white,draw=lightgray,pos=0.5] (18);
\draw [color=lightgray] (a6) -- (a8) coordinate[cs1,fill=white,draw=lightgray,pos=0.5]  (8);
\draw [color=lightgray] (a7) -- (a8) coordinate[cs1,fill=white,draw=lightgray,pos=0.5] (12);

\draw [lightgray] plot [smooth] coordinates {(a1) (aux1) (M) (O)};
\draw [lightgray] plot [smooth] coordinates {(a8) (aux2) (M) (N)};


\draw (a1) coordinate[c2,fill=red,label=below:$a_1$];

\draw (a2) coordinate[c1,fill=lightgray,label=left:$a_2$];

\draw (a3) coordinate[c1,fill=lightgray,label=above:$a_3$];

\draw (a4) coordinate[c1,fill=lightgray,label=left:$a_4$];

\draw (a5) coordinate[c1,fill=lightgray,label=right:$a_5$];

\draw (a6) coordinate[c1,fill=lightgray,label=left:$a_6$];

\draw (a7) coordinate[c1,fill=lightgray,label=above:$a_7$];

\draw (a8) coordinate[c2,draw=green,fill=white,label=above:$a_8$];

\draw (N) coordinate[c2,fill=red,label=right:$N$];

\draw (M) coordinate[c2,draw=green,fill=white,label=right:$M$];

\draw (O) coordinate[c2,draw=green,fill=white,label=right:{\scriptsize \color{lightgray}$O$}];

\end{tikzpicture}
}
\\
(c) & (d)
\end{tabular}
\end{center}
\caption{\label{2020-hardy-TITS}
Orthogonality hypergraphs of
(a) the Hardy gadget extended to a true-implies-true gadget~\cite{2018-minimalYIYS}, as enumerated in (c),
with 8+2=10 contexts and 21+4=25 atoms
$\{
 \{dd,8,9,cv\}     $,
$\{dd,11,12,vc\}   $,
$\{cv,vu,uu,dv\}   $,
$\{vc,uv,uu,vd\}   $,
$\{vu,18,19,uv\}   $,
$\{vd,2,3,\Psi \}  $,
$\{uu,20,21,\Psi \}$,
$\{dv,16,17,\Psi \}$,
$\{5,dd,M,N \}$,
$\{1,O,P,M \}
\}$;
(b) rendition based on the true-implies-true extended Specker bug or cat's cradle gadget~\cite{kochen1,svozil-2017-b}, as enumerated in (c),
with 7+2=9 contexts and 13+3=17 atoms
$\{
\{a_8,.,a_6\}  $,
$\{a_8,.,a_7\}  $,
$\{a_6,a_4,a_3\}  $,
$\{a_7,a_5,a_2\}  $,
$\{a_4,.,a_5\}    $,
$\{a_2,.,a_1 \}   $,
$\{a_3,.,a_1 \} $,
$\{a_8,M,N \} $,
$\{a_1,.,M \}
\}$.
Due to the way these true-implies-false gadgets are constructed $N$ always turns out to be~1 if $\Psi$ or $a_1$ are supposed to be~1.
Small circles indicate ``auxiliary'' observables which can be chosen freely,
subject to orthogonality constraints: all smooth lines indicate respective contexts representing orthonormal bases.}
\end{figure}
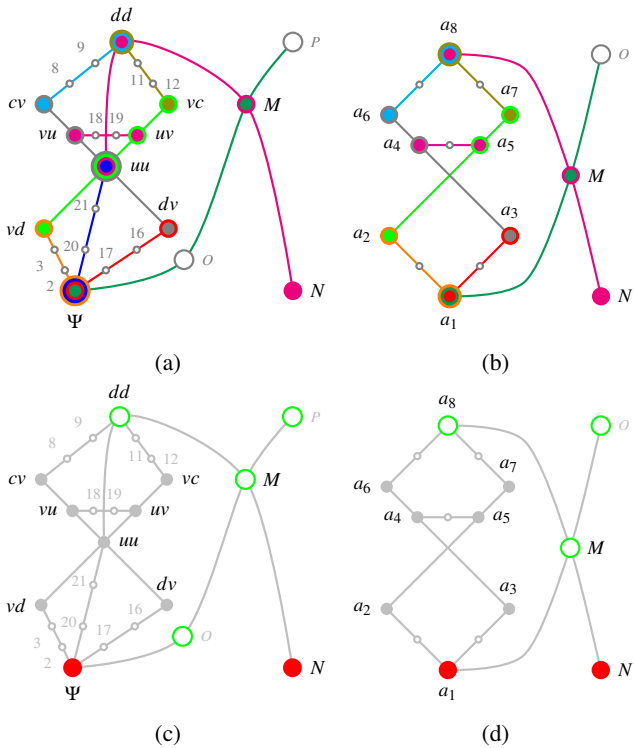

Figure~\ref{2020-hardy-TITS} depicts the extension of the Hardy gadget which delivers a classical true-implies-true prediction at its terminal points $\Psi$ and $N$.
A faithful orthogonal representation of the extended Hardy gadget can be obtained {\it ad hoc}
by the heuristic algorithm {\tt VECFIND}~\cite{Pavii2018}
in the coordinate basis $\{0,\pm 1,\pm 2,3\}$ and the {\tt -nk} option, which
is capable of finding ``almost all'' vectors,
including the true-implies-true terminal points $\Psi$ and $N$ {\it ex machina}, and (for this coordinate basis) needs a little helping hand
(or the additional component basis elements $\{-3,5,7, \sin \theta , \cos \theta $)
with $\theta \neq n\pi/4$, $n\in \mathbb{Z}$ to find the
complete set, given by
$\Psi=\begin{pmatrix}0,1,1,-1\end{pmatrix}$,
$2=\begin{pmatrix}2,2,-1,1\end{pmatrix}$,
$3=\begin{pmatrix}3,-2,1,-1\end{pmatrix}$,
$vd=\begin{pmatrix}0,0,1,1\end{pmatrix}$,
$uu=\begin{pmatrix}1,0,0,0\end{pmatrix}$,
$vu=\begin{pmatrix}0,0,0,1\end{pmatrix}$,
$cv=\begin{pmatrix}0,1,1,0\end{pmatrix}$,
$8=\begin{pmatrix}3,1,-1,2\end{pmatrix}$,
$9=\begin{pmatrix}-2,1,-1,2\end{pmatrix}$,
$dd=\begin{pmatrix}0,-1,1,1\end{pmatrix}$,
$11=\begin{pmatrix}3,-2,-1,-1\end{pmatrix}$,
$12=\begin{pmatrix}2,2,1,1\end{pmatrix}$,
$vc=\begin{pmatrix}0,0,1,-1\end{pmatrix}$,
$uv=\begin{pmatrix}0,1,0,0\end{pmatrix}$,
$dv=\begin{pmatrix}0,1,-1,0\end{pmatrix}$,
$16=\begin{pmatrix}-2,1,1,2\end{pmatrix}$,
$17=\begin{pmatrix}3,1,1,2\end{pmatrix}$,
$18=\begin{pmatrix}\cos \theta , 0, \sin \theta , 0\end{pmatrix}$,
$19=\begin{pmatrix}-\sin \theta  , 0, \cos \theta , 0\end{pmatrix}$,
$20= \begin{pmatrix}0, 4, -3, 1\end{pmatrix}$,   
$21=\begin{pmatrix}0,2,5,7\end{pmatrix}$,
$M=\begin{pmatrix}0,1,0,1\end{pmatrix}$,
$N=\begin{pmatrix}0,1,2,-1\end{pmatrix}$,
$O=\begin{pmatrix}2,-1,2,1\end{pmatrix}$,
$P=\begin{pmatrix}3,1,-2,-1\end{pmatrix}$,
where $\theta \neq n\pi/4$, $n\in \mathbb{Z}$.
[The original coordinatization suggested for atom $20$ was
$\begin{pmatrix}0,1,-1,0\end{pmatrix}$ but a completion  would have resulted in duplicities,
namely, $21=\begin{pmatrix}0,1,1,2\end{pmatrix}=dv$; and therefore the original suggestion had to be dropped.]
Although we do not concentrate on maximal violations of classical predictions by quantum probabilities for reasons mentioned later,
it is worth noting that, as $\vert \langle \psi \vert N \rangle \vert^2 = 8/9$, the quantum violation of the classical predictions will, in this particular configuration,
occur in one out of nine times; that is, with probability~$0.\overline{1}$ (proper normalization is always assumed).
Fortuitously, if one concentrates on the quantum signal for observable $\vert dd\rangle \langle dd \vert$, then one obtains the same
quantum prediction  $\vert \langle \psi \vert dd \rangle \vert^2 = 1/9$ for this outcome---although classically it should never occur.


\section{Extensions to gadgets with indistinguishable classical truth assignments}

One way to proceed would be what Kochen and Specker did with their true-implies-true gadget $\Gamma_1$, and serially compose them at their respective
(properly parametrized)
terminal points often enough to obtain $\Gamma_2$, which renders a complete contradiction with exclusivity~\cite{kochen1}.
Instead of this head-on strategy for obtaining complete contradictions with classical noncontextual hidden variable models
we shall use a more subtle approach and consider a hypergraph which, again in analogy with Kochen and Specker's $\Gamma_3$ in three dimensions,
cannot be classically embedded in a Boolean algebra.
The construction uses two true-implies-true extended Hardy gadgets to construct two pairs of observable propositions which cannot be
differentiated by classical two-valued measures---and thus by any classical probability distributions----although ``plenty''
such two-valued states still exist
(but their set is ``too meager'' to allow mutual distinguishability of all pairs of distinct propositions).

\begin{figure*}
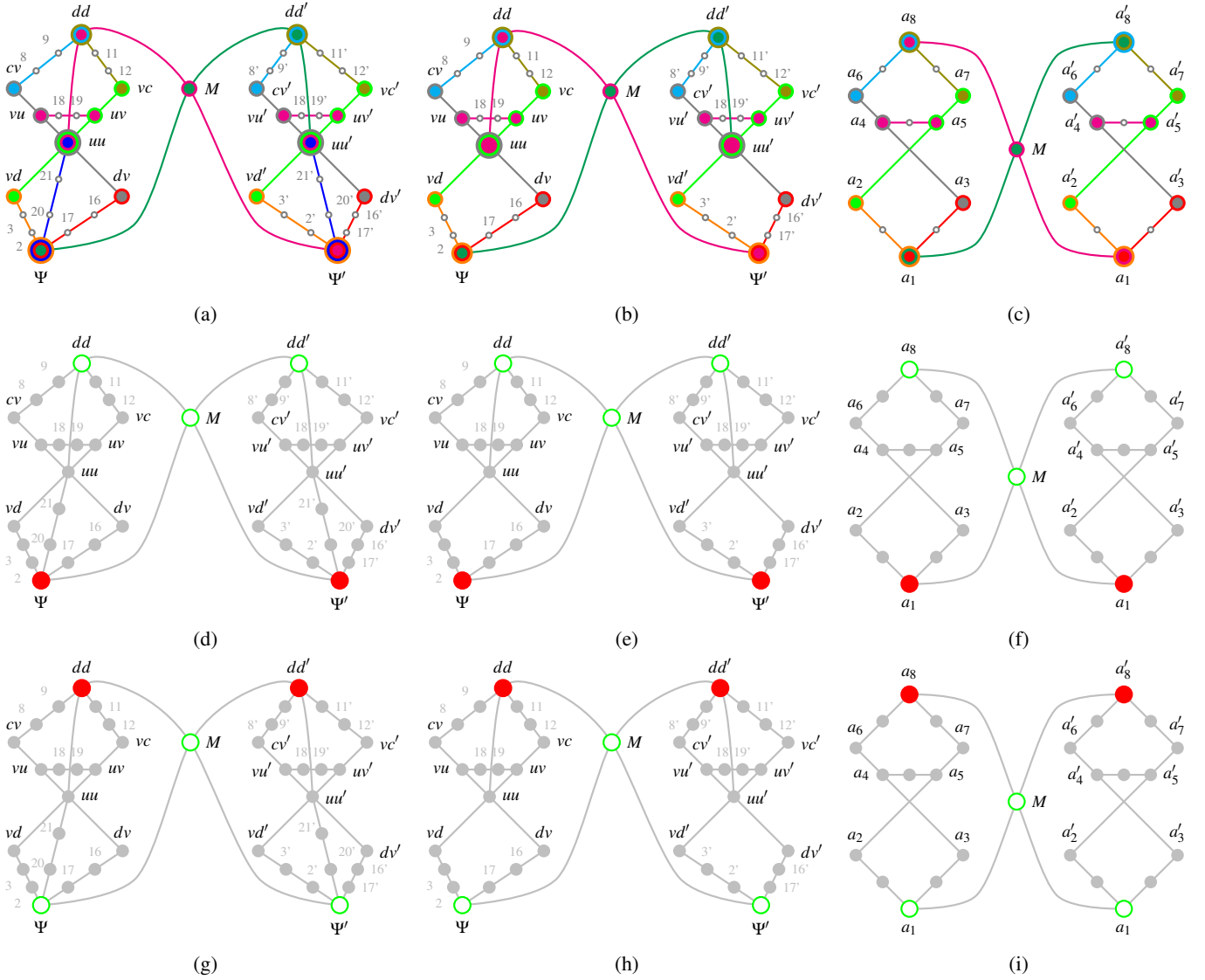

\begin{center}

\end{center}
\caption{\label{2020-hardy-TIFFT}
Orthogonality hypergraphs of
(a) a combination of two extended Hardy gadgets form a structure of quantum observables which cannot be classically embedded
with $2 \times 8+2=18$ contexts and $2 \times 21+1=43$ atoms that cannot be classically embedded
because of the indistinguishability with classical means (two-valued states) of the two pairs of atoms $\{\psi_1,\psi_2\}$ as well as $\{dd_1,dd_2\}$, respectively;
(b) a combination of two extended Hardy-like gadgets first introduced in Fig.~4(b) of Ref.~\cite{2018-minimalYIYS} form a structure of quantum observables
with $2 \times 7=14$ contexts and $2 \times 19 +1 =39$ atoms that cannot be classically embedded
because of the indistinguishability with classical means (two-valued states) of the two pairs of atoms $\{\psi_1,\psi_2\}$ as well as $\{dd_1,dd_2\}$, respectively;
(c) a combination of two extended Specker bug/cat's cradle gadgets~\cite{kochen1,svozil-2017-b}
with $2 \times 7+2=16$ contexts and $2 \times 13+1 = 27$ atoms that cannot be classically embedded
because of the indistinguishability with classical means (two-valued states) of the two pairs of atoms $\{a_1,a_1'\}$ as well as $\{a_8,a_8'\}$, respectively.
(d)---(i) depict the associated two-valued states
which are not 0 on all four observables $\{\psi_1,\psi_2,dd_1,dd_2\}$ as well as $\{a_1,a_1',a_8,a_8'\}$, respectively.
Only valuations that are relevant for the proof are drawn.
Small circles indicate ``auxiliary'' observables which can be choosen freely,
subject to orthogonality constraints: all smooth lines indicate respective contexts representing orthonormal bases.}
\end{figure*}

\begin{figure}
\begin{center}
\begin{tabular}{ c c }
\resizebox{.193\textwidth}{!}{%
\begin{tikzpicture}  [scale=1.2]

\tikzstyle{every path}=[line width=1pt]

\newdimen\ms
\ms=0.1cm
\tikzstyle{s1}=[color=red,rectangle,inner sep=3.5]
\tikzstyle{c4}=[circle,inner sep={\ms/8},minimum size=5*\ms]
\tikzstyle{c3}=[circle,inner sep={\ms/8},minimum size=4*\ms]
\tikzstyle{c2}=[circle,inner sep={\ms/8},minimum size=3*\ms]
\tikzstyle{c1}=[circle,inner sep={\ms/8},minimum size=2*\ms]
\tikzstyle{cs1}=[circle,inner sep={\ms/8},minimum size=1*\ms]


\coordinate (psi1) at (0,0);
\coordinate (psi2) at (2,0);
\coordinate (M) at (1,1);
\coordinate (dd1) at (0.5,1.5);
\coordinate (dd2) at (1.5,1.5);
\coordinate (uu1) at (0,2);
\coordinate (uu2) at (2,2);


\draw [color=orange] (psi1) -- (uu2);
\draw [color=blue] (psi2) -- (uu1);
\draw [color=red] (psi1) -- (uu1)  coordinate[c1,fill=white,draw=lightgray,pos=0.33,label=left:{\color{lightgray}{$20$}}] (20)   coordinate[c1,fill=white,draw=lightgray,pos=0.66,label=left:{\color{lightgray}{$21$}}] (21);
\draw [color=green] (psi2) -- (uu2) coordinate[c1,fill=white,draw=lightgray,pos=0.33,label=right:{\color{lightgray}{$20'$}}] (202)   coordinate[c1,fill=white,draw=lightgray,pos=0.66,label=right:{\color{lightgray}{$21'$}}] (212);    ;


\draw (psi1) coordinate[c2,fill=orange,label=below:$\Psi$];
\draw (psi1) coordinate[c1,fill=red];

\draw (psi2) coordinate[c2,fill=green,label=below:$\Psi'$];
\draw (psi2) coordinate[c1,fill=blue];

\draw (M) coordinate[c2,fill=orange,label=below:$M$];
\draw (M) coordinate[c1,fill=blue];

\draw (dd1) coordinate[c2,fill=blue,label=above:$dd$];

\draw (dd2) coordinate[c2,fill=orange,label=above:$dd'$];

\draw (uu1) coordinate[c2,fill=blue,label=above:$uu$];
\draw (uu1) coordinate[c1,fill=red];

\draw (uu2) coordinate[c2,fill=orange,label=above:$uu'$];
\draw (uu2) coordinate[c1,fill=green];

\end{tikzpicture}
}
$\quad$
&
$\quad$
\resizebox{.15\textwidth}{!}{%
\begin{tikzpicture}  [scale=1.2]

\tikzstyle{every path}=[line width=1pt]

\newdimen\ms
\ms=0.1cm
\tikzstyle{s1}=[color=red,rectangle,inner sep=3.5]
\tikzstyle{c4}=[circle,inner sep={\ms/8},minimum size=5*\ms]
\tikzstyle{c3}=[circle,inner sep={\ms/8},minimum size=4*\ms]
\tikzstyle{c2}=[circle,inner sep={\ms/8},minimum size=3*\ms]
\tikzstyle{c1}=[circle,inner sep={\ms/8},minimum size=2*\ms]
\tikzstyle{cs1}=[circle,inner sep={\ms/8},minimum size=1*\ms]


\coordinate (psi1) at (0,0);
\coordinate (psi2) at (2,0);
\coordinate (M) at (1,1);
\coordinate (dd1) at (0.5,1.5);
\coordinate (dd2) at (1.5,1.5);
\coordinate (uu1) at (0,2);
\coordinate (uu2) at (2,2);


\draw [color=orange] (psi1) -- (uu2);
\draw [color=blue] (psi2) -- (uu1);


\draw (psi1) coordinate[c2,fill=orange,label=below:$\Psi$];

\draw (psi2) coordinate[c2,fill=blue,label=below:$\Psi'$];

\draw (M) coordinate[c2,fill=orange,label=below:$M$];
\draw (M) coordinate[c1,fill=blue];

\draw (dd1) coordinate[c2,fill=blue,label=above:$dd$];

\draw (dd2) coordinate[c2,fill=orange,label=above:$dd'$];

\draw (uu1) coordinate[c2,fill=blue,label=above:$uu$];

\draw (uu2) coordinate[c2,fill=orange,label=above:$uu'$];

\end{tikzpicture}
}
\\
(a)&(b)
\end{tabular}
\end{center}
\caption{\label{2020-hardy-TIFFTbb}
Hypergraphs of the ``orthogonality backbones'' of (a) Fig.~\ref{2020-hardy-TIFFT}(a), and (b)
Fig.~\ref{2020-hardy-TIFFT}(b) supporting the two-valued states depicted in Fig.~\ref{2020-hardy-TIFFT}(c)
and (d), respectively.
}
\end{figure}
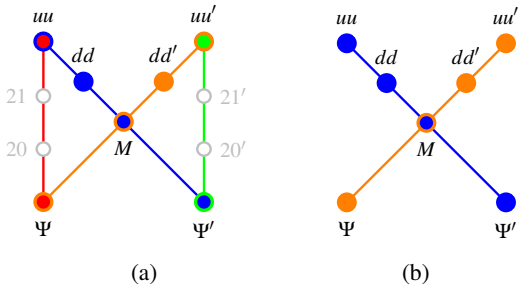

Searches for a faithful orthogonal representation of an extensions
using two (a combination) of the ``original'' version of the Hardy gadget, as depicted in Fig.~\ref{2020-hardy-TIFFTbb}(a),
have been inconclusive so far.
Nevertheless, as it turns out this task can be completed by using (a combination of)
a slight modification of Hardy's gadget introduced in Fig.~4(b) of Ref.~\cite{2018-minimalYIYS},
in which the original context $\{ \Psi , .  .  ,uu \}$ is ``relocated''
or ``reshuffled'' into the context $\{ uu ,  .  .  ,dd \}$.
The resulting gadget, depicted in Fig.~\ref{2020-hardy-TIFFT}(b), not only has less atoms but, most importantly,
has a less tight ``orthogonality backbone'' structure, depicted in Fig.~\ref{2020-hardy-TIFFTbb}(b),
of just two contexts intertwined in a single atom $M$, namely
$\{
\{uu_1,dd_1,M,\Psi_2\}$,
$\{uu_2,dd_2,M,\Psi_1\}
\}$,
as compared to the tight configuration resulting from a composition of two of Hardy's original gadgets
$\{
\{Psi_1, .  .  , uu_1\}$,
$\{uu_1,dd_1,M,\Psi_2\}$,
$\{uu_2,dd_2,M,\Psi_1\}$,
$\{Psi_2, .  .  , uu_2\}
\}$ depicted in Fig.~\ref{2020-hardy-TIFFTbb}(a).

More explicitly, {\tt VECFIND}~\cite{Pavii2018},
with the component basis
$\{
0,\pm 1,2,-3,4,5
\}$,
yields an ad hoc coordinatization
of the intertwine atoms
$\psi=\begin{pmatrix}1,0,0,0\end{pmatrix}$,
$vd=\begin{pmatrix}0,2,-1,1\end{pmatrix}$,
$uu=\begin{pmatrix}0,0,1,1\end{pmatrix}$,
$vu=\begin{pmatrix}1,-1,1,-1\end{pmatrix}$,
$cv=\begin{pmatrix}-3,-1,1,-1\end{pmatrix}$,
$dd=\begin{pmatrix}1,-3,0,0\end{pmatrix}$,
$vc=\begin{pmatrix}-3,-1,-1,1\end{pmatrix}$,
$uv=\begin{pmatrix}1,-1,-1,1\end{pmatrix}$,
$dv=\begin{pmatrix}0,2,1,-1\end{pmatrix}$,
$\psi'=\begin{pmatrix}-3,-1,0,0\end{pmatrix}$,
$vd'=\begin{pmatrix}1,-3,4,-1\end{pmatrix}$,
$uu'=\begin{pmatrix}0,1,1,1\end{pmatrix}$,
$vu'=\begin{pmatrix}-3,-1,2,-1\end{pmatrix}$,
$cv'=\begin{pmatrix}1,0,1,-1\end{pmatrix}$,
$dd'=\begin{pmatrix}0,2,-1,-1\end{pmatrix}$,
$vc'=\begin{pmatrix}5,0,-1,1\end{pmatrix}$,
$uv'=\begin{pmatrix}-1,-3,-1,4\end{pmatrix}$,
$dv'=\begin{pmatrix}1,-3,1,2\end{pmatrix}$, and
$M=\begin{pmatrix}0,0,1,-1\end{pmatrix}$
which can be readily completed into a faithful orthogonal representation of
the hypergraph depicted in Fig.~\ref{2020-hardy-TIFFT}(b).
Note that in this particular configuration, because of indistinguishability,
the classical prediction to find a particle prepared in a state $\Psi$
in the state $\Psi'$ is one (certainty), whereas quantum mechanics predicts nonoccurrence of
the elementary propositional observable $\vert \Psi' \rangle  \langle \Psi'\vert$
given a preselected, prepared state $\vert \Psi \rangle$
with probability
$\vert \langle \Psi \vert \Psi' \rangle \vert^2 = 9/10$; that is,  the violation
of the classical prediction by quantum mechanics occurs in this case in one out of ten experimental runs.

\section{Summary and cautionary remarks}

Hardy-type configurations have been extended to configurations of contexts which show a different nonclassical performance:
they contain distinct quantum observables that cannot be distinguished from one another by any classical (noncontextual) means.
To appreciate the difference of this aspect beyond the realization of just another relational property
among some prepared state and its measurement, it is important to keep in mind that,
according to a finding by Kochen and Specker~\cite[Theorem-0]{kochen1},
indistinguishability serves a demarcation criterion for strong forms of nonclassicality:
The absence of classical distinguishability indicates a stronger contraindication of hidden-variable theories (relative to the assumptions)
than, say, exploitation of true-implies-\{true,false\} properties~\cite{2018-minimalYIYS} which still allow faithful classical
embeddability of the quantum observables (albeit with different statistical predictions), and merely requires complementarity.

Indistinguishability by classical means indicates a rather strong form of nonclassicality -- that is,
the impossibility to faithfully embed the quantum mechanical observables in classical Boolean structures --
while still allowing the direct experimental falsification of the respective quantum and classical predictions.
Thereefore, it is not affected by questions related to the empirical pertinence of Kochen-Specker proofs (by contradiction) of the
absence of any classical interpretation, stated pointedly by Clifton (quoted from my memory~\cite{clifton}),
``how can you measure a contradiction?''

Further efforts could advance by ``improving the nonclassical performance'' of gadgets
not in terms of the number of (counterfactual, complementary) observables
but in terms of quantum-to-classical discords in four dimensions.
The hypergraph method developed earlier might suggest such advancements by their emphasis
on the logico-algebraic structure, thereby making possible a more systematic explotation of
feasible configurations of observables.
There already exist true-implies-\{true,false\} gadgets which
yield high performance in three
dimensions~\cite{svozil-2018-whycontexts,Svozil-2018-p,Ramanathan-18}.

Nevertheless, once the vectors corresponding to pre- and postselected states are fixed,
it is always possible to find any kind of conforming or disagreeing classical-versus-quantum behavior.
As I have pointed out elsewhere~\cite{svozil-2020-c}, these kind of statements are contingent on the chosen gadget consisting of mostly counterfactual observables ``in the mind''
of the observer~\cite{berkeley,stace}.
Nevertheless, any such considerations raise fascinating,
challenging issues in a variety of fields which might have been perceived unrelated so far:
graph theory, (linear) algebra, functional analysis, geometry, automated theorem proving and---last but not least---quantum physics and
quantum information (processing) technology.

\begin{acknowledgments}
This research was funded in whole, or in part, by the Austrian Science Fund (FWF), Project No. I 4579-N.

The author declares no conflict of interest.

I gratefully acknowledge enlightening discussions with Adan Cabello, Hans Havlicek, Norman Megill, Mladen Pavi{\v{c}}i{\'{c}}, Jos\'{e} R. Portillo, and Mohammad Hadi Shekarriz.
Christian Jendreiko has identified an omission of three atoms (atoms 19, 20, and 21) in the partition logic representation of Table~\ref{2020-hardy-tablepartitionlogic} in an earlier version of the manuscript.
I am grateful to Josef Tkadlec for providing a {\em Pascal} program which computes and analyses the set of two-valued states of collections of contexts.
I am also grateful to  Norman D. Megill and Mladen Pavi{\v{c}}i{\'{c}} for providing a {\em C++} program which computes the faithful orthogonal representations of hypergraphs written in MMP format, given possible vector components.
All misconceptions and errors are mine.
\end{acknowledgments}

\end{document}